\documentclass[acmsmall]{acmart}
\usepackage{colortbl}
\usepackage{multicol}
\usepackage{multirow}
\usepackage{booktabs}
\usepackage{graphicx}
\usepackage{caption}
\usepackage{subcaption}
\usepackage{hhline}
\usepackage{xcolor}
\usepackage{soul}
\newcommand{\hlc}[2][white]{{%
    \colorlet{foo}{#1}%
    \sethlcolor{foo}\hl{#2}}%
}

\AtBeginDocument{%
  \providecommand\BibTeX{{%
    \normalfont B\kern-0.5em{\scshape i\kern-0.25em b}\kern-0.8em\TeX}}}

\setcopyright{acmcopyright}
\acmJournal{PACMHCI}
\acmYear{2022} \acmVolume{6} \acmNumber{CSCW2}
\acmArticle{528} \acmMonth{11} \acmPrice{15.00}
\acmDOI{10.1145/3555641}

\usepackage{mathtools}
\usepackage{algorithm}
\usepackage{algpseudocode}

\begin{document}

\title{Multi-Objective Personalization in Multi-Stakeholder Organizational Bulk E-mail:  A Field Experiment}

\author{Ruoyan Kong}
\email{kong0135@umn.edu}
\affiliation{%
  \institution{University of Minnesota - Twin Cities}
  \country{USA}
}
\author{Charles Chuankai Zhang}
\affiliation{%
  \institution{University of Minnesota - Twin Cities}
  \country{USA}
}
\author{Ruixuan Sun}
\affiliation{%
  \institution{University of Minnesota - Twin Cities}
  \country{USA}
}
\author{Vishnu Chhabra}
\affiliation{%
  \institution{University of Minnesota - Twin Cities}
  \country{USA}
}
\author{Tanushsrisai Nadimpalli}
\affiliation{%
  \institution{University of Minnesota - Twin Cities}
  \country{USA}
}
\author{Joseph A. Konstan}
\email{konstan@umn.edu}
\affiliation{%
  \institution{University of Minnesota - Twin Cities}
  \country{USA}
}

\renewcommand{\shortauthors}{Ruoyan Kong et al.}

\begin{abstract}
Bulk email is often used in organizations to communicate ``important-to-organization'' messages such as policy changes, organizational plans, and administrative updates. However, normal employees may prefer messages more relevant to their jobs or interests. Organizations face the challenge of balancing prioritizing the messages they prefer employees to know (tactical goals) while maintaining employees' positive experiences with these bulk emails, then they continue to read these emails in the future (strategic goals).

Could personalization help organizations achieve these tactical and strategic goals? In an 8-week field experiment with a university newsletter, we implemented a 4x5x5 factorial design on personalizing subject lines, top news, and message order based on both the employees' and the organization's preferences. We measured these designs' influences on the open/interest/recognition/read-in-detail rate of the whole newsletter and the single messages within it.

We found that ``important-to-organization'' messages only got higher recognition rates when being put on subject lines / top news (tactical goal). Mixing them with employee-preferred messages in top news did not bring further improvement to their own recognition rates but could improve the whole newsletter's recognition rate. Only when the top news solely contained the employee-preferred messages were the employees slightly more interested in the newsletter (strategic goal). We further analyze on which topics the employees and the organization's preferences conflicted. Finally, we discuss the design suggestions for organizational bulk email.\footnote{This is a pre-print version of a paper accepted to CSCW 2022 — The 25th ACM Conference On Computer-Supported Cooperative Work And Social Computing.}
\end{abstract}

\ccsdesc[500]{Human-centered computing~Empirical studies in collaborative and social computing}

\keywords{organizational communication, email, personalization, field experiment}

\maketitle

\section{Introduction}
Organizations often send \textbf{bulk emails} (e.g., newsletters, all-company emails) to employees \cite{10.1145/1180875.1180941} to announce events, policy changes, and administrative updates. In the university we conducted this study with, an employee receives 30 bulk emails per week from the central offices on average. Many of them contain tens of \textbf{single messages}, which means make each employee receives over 250 single pieces of information from these bulk emails per week. Figure \ref{fig:brief} is an example university bulk email \hlc[white]{of our study site University of Minnesota: U of M Brief}. It is a weekly newsletter sent to all the employees across all 5 campuses with around 30 messages and 7 sections (top news, u-wide news, and each campus' news) in each newsletter.

\begin{table}
\begin{minipage}{.48\textwidth}

\centering
  \includegraphics[width=1\columnwidth]{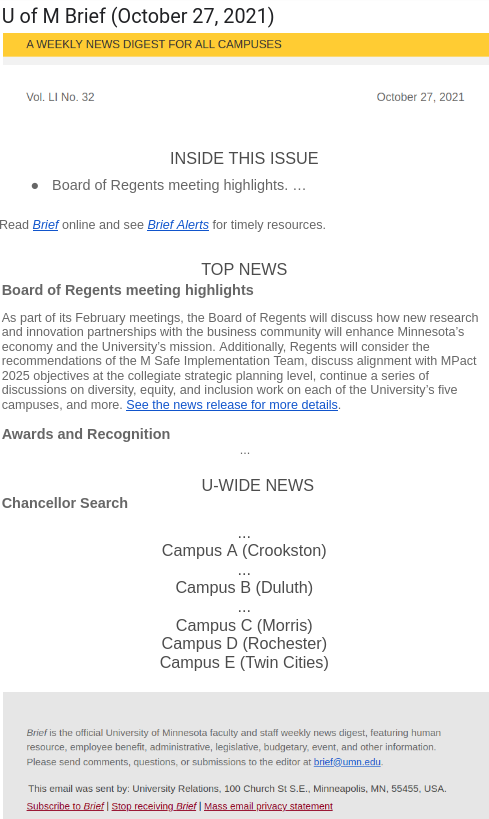}
  \captionof{figure}{A sample Brief on Oct 27, 2021. The messages within each section are hidden in this picture.}~\label{fig:brief}

\end{minipage}%
\hspace{0.05in}
\begin{minipage}{.46\textwidth}
\centering
  \includegraphics[width=1\columnwidth]{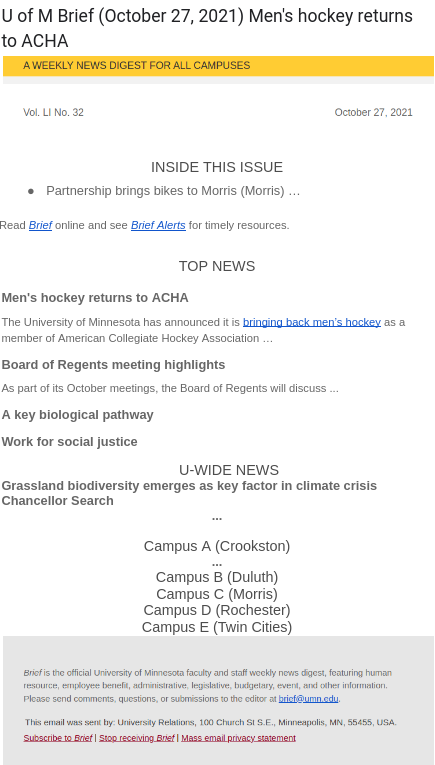}
  \captionof{figure}{A sample personalized Brief. The messages within each section are hidden in this picture.}~\label{fig:per_brief}
\end{minipage}%
\end{table}

\hlc[white]{These bulk emails (\textbf{organizational bulk email}) have \textbf{multiple stakeholders}, including the information sources (e.g., organizations' leaders), the \textbf{communicators} (the communication staff in charge of designing and sending bulk emails, e.g., newsletter editors), the recipients (the employees who receive these bulk emails), managers, etc. In this paper, we focus on 2 kinds of stakeholders: the organization leaders and communicators who represent the organization's perspective, and the normal employees who represent the recipients' perspective on messages' work-relevance/interest/importance level. The organization and employees may have different perspectives \mbox{\cite{welch2012appropriateness, kong2021learning, kang2020organizational}}. For example, the university's board of regents might want all the employees to know about their meeting updates; then they put it as the first message of the Brief with the hope that all the employees who opened this Brief will notice this message (see Figure \mbox{\ref{fig:brief}}). However, the employees might perceive it as unactionable \textbf{high-level information} and view this Brief as irrelevant to their job. They might stop reading Brief after seeing this message.}

\hlc[white]{Therefore, designing organizational bulk emails is a \textbf{multi-objective problem} for organizations. Their \textbf{tactical} objective is to use bulk emails to make employees aware of ``important'' messages, such as the board of regents meeting updates. At the same time, they have the \textbf{strategic} objective of maintaining the effectiveness of their communication channels by ensuring employees see messages they perceive as relevant and continue to read in the future. Organizations need to balance their short-term tactical goals and long-term strategic goals in designing these emails.}

\hlc[white]{Here we see the opportunity of exploring personalization based on both \textbf{organization's preference} (organization's view on message's work-relevance and importance to employees as assessed by the newsletter editors) and \textbf{employee's preference} (employee's view on message's work-relevance and interest level to themselves) to help organizations reach these communication goals. For example, a design we tried with the newsletter above (see Figure \mbox{\ref{fig:per_brief}}) is to put \textbf{organization-preferred messages} (e.g., board of regents meetings) near to \textbf{employee-preferred messages} (e.g., Men's hockey, in a case where the specific employee like sports) to balance these two interests. Specifically, we conducted an 8-week field experiment with a university newsletter and 141 employees. We employed a 4x5x5 factorial design in personalizing subject line/top news/message order. We measured these designs' influences on the open/interest/recognition rate of the whole newsletter (strategic goals) and the recognition/read-in-detail rate of the messages within it (tactical goals).}

 Regarding the tactical goals, we found that ``important-to-organization'' messages could only get higher recognition rates but not higher read-in-detail rates when being put on subject lines / top news. Mixing them with employee-preferred messages in top news did not bring further improvement to their own recognition rates but brought better performance on the strategic goals --- they improved the whole newsletter's recognition rate. However, only when the top news solely contained the employee-preferred messages would the employees be marginally more interested in the newsletter. We looked into the organization's preferences and the employees' preferences on message topics and found that the conflicts over what is important/relevant existed widely.
 
\hlc[white]{Our contributions include two parts. Regarding the fundamental theoretical advances, we studied the unique personalization problem with organizational bulk emails where the short-term tactical goals might not align with the longer-term strategic goals, while commercial bulk emails' tactical goals (e.g., recipients purchase the recommended products) and strategic goals (e.g., recipients perceive their emails as useful and keep reading them next time) are aligned \mbox{\cite{sahni2018personalization, wattal2012s, singh2015email}}. Also, we conducted an 8-week field experiment to enable employees to get used to the experimental newsletters, while the research on organizational emails is mainly lab experiments \mbox{\cite{10.1145/3290605.3300604}}, dataset analysis (Enron and Avocado \mbox{\cite{10.1145/3077136.3080782, klimt2004enron, bermejo2011improving}}), and observational field studies \mbox{\cite{10.1145/2858036.2858262, 10.1145/859670.859673, paczkowski2016checking, kong2021virtual}}. Regarding the practical advances, we provided the tradeoffs and suggestions for organizations in designing bulk emails. We designed a personalization framework for organization's bulk emails, including the process to collect stakeholders' preferences, the algorithms to generate personalized bulk emails, and the mechanisms to evaluate their performance. The rest of this paper includes related work (2), background (3), methods (4), results (5), and discussion (6). This study was approved by the IRB of the University of Minnesota.}

\section{Related Work \& Gaps}
\subsection{Organizational Communication and Bulk Email's Goals}
Organizational communication has been defined by various disciplines \cite{myers1982managing, katz2008communication}. Wrench and Punyanunt-Carter \cite{wrench2012introduction} define it as: \textit{``the process whereby an organizational
stakeholder(s) attempts to stimulate meaning in the mind of another organizational stakeholder(s) through the intentional use of verbal, nonverbal, and mediated messages.''} Organizational communication is a problem with multiple stakeholders, including the information providers, the information gatekeepers, the information recipients, and the organization itself. Instead of prioritizing one single stakeholder's preferences, the goal of organizational communication is to maintain the whole organization's productivity \cite{welch2007rethinking} by developing the awareness of organizational tasks, promoting a positive sense of belonging, and developing the understanding of organizational needs and goals. 

Several studies have shown that internal organization emails, especially the organizational bulk email system, usually fail the communication goals above. First, employees are often too overloaded to have a good recognition level of or experience with these emails. Dabbish and Kraut \cite{10.1145/1180875.1180941} conducted a nationwide organization survey and found that email volume is positively correlated with the feeling of email overload at work in 2006. Grevet et al.\cite{grevet2014overload} interviewed 19 Gmail users and found the problem of email overload became more serious in 2014 as the size of inbox and the number of unread messages increased hugely compared to Whittaker and Sidner's findings in 1996 \cite{whittaker1996email}. Second, the stakeholders naturally have different preferences on the information, which stops them from paying attention to these emails. Dabbish et al. surveyed 128 employees of a university and found that employees prioritized the emails from the senders that they had a direct work relationship with \cite{dabbish2005understanding}. Kong et al. \cite{kong2021learning} interviewed the recipients, managers, and communicators of a university. They found that the managers and communicators thought that employees should know about what was going on in the university. But the employees felt that these bulk emails were unactionable and irrelevant. Then they questioned the credibility of these bulk emails and gradually decided to stop reading them. 



\subsection{Bulk Email Personalization}

\subsubsection{Bulk Email Personalization's Content}
Personalization has been pointed as a solution for email overload \cite{cecchinato2014personalised} and several studies used personalization to improve commercial bulk emails' performance. The personalization content could be demographic information like names \cite{sahni2018personalization, wattal2012s}, majors, departments \cite{trespalacios2016effects}; or preference information like browsing history \cite{wattal2012s}, deals or tools recommendation \cite{singh2015email, hawkins2008understanding}. Though Sahni et al.'s experiments found adding recipients' names to subject lines useful \cite{sahni2018personalization}, many other studies supported that the personalization based on preferences performed better than the personalization based on demographics. Wattal et al. \cite{wattal2012s} found that customers responded negatively to emails with identifiable information. Trespalacios and Perkins also found the effect of adding identifiable demographic information insignificant in an experiment with a university email \cite{trespalacios2016effects}. Hawkins et al. pointed out that personalized messages need to provide the recipients with new information about themselves instead of simply adding names or addresses \cite{hawkins2008understanding}. Wattal et al. \cite{wattal2012s} personalized email content based on customers' purchasing preferences and received positive responses. We personalize based on preferences instead of demographics in this study. Because we could not simply add the recipients' names to every internal newsletter of the organization --- the employees could quickly learn that seeing their names in the organizational bulk emails means nothing special.

\subsubsection{Bulk Email Personalization's Design}
In this section, we looked at the categories of the personalization designs of commercial bulk emails. Besides that, we also searched the work around the advertising placements of web pages as we considered a similarity between attracting users' attention to advertisements and attracting employees' attention to the ``important-to-organization'' bulk messages.
\begin{enumerate}
\item Subject line: An informative subject line was believed to be a key factor in successful email marketing \cite{waldow2012rebel}. Sahni et al. added recipients' names to subject lines \cite{sahni2018personalization}, and that method increased a marketing email's open rate by 20\%. However, there are also studies showing that an uninformative subject line could create an information gap that attracts recipients to open emails \cite{doi:10.1080/13645579.2015.1078596,callegaro2009effect}. The ``Long vs. Short Email Subject Line Test'' of WhichTestWon.com in 2011 \cite{waldow2012rebel} found that a longer subject line led to a higher open rate. But Alchemy Worx's test on a discount promotion email found that a longer subject line led to lower open rates \cite{worx2016subject}. An explanation for the contradictory results is that the longer subject lines only influence by providing more information. Under this theory, the factor that actually matters here is whether the topic of the subject line matches with the recipients' preferences. The longer subject lines might get lower open rates but higher action rates, because only those who are interested in it will open it \cite{worx2016subject,jaidka2018predicting}. 

\item Top section: The traditional theory is that users would pay more attention to the top positions during browsing \cite{shrestha2007eye}. Wattal et al. \cite{wattal2012s}, and Trespalacios and Perkins \cite{trespalacios2016effects} tried adding recipients' names, majors, or departments to the greetings or the first paragraphs of emails in their studies and Wattal et al. found that greetings influenced their customers' response rates significantly. 

\item Selection of contents: Many studies personalized commercial bulk emails by selecting the most interesting content for recipients. Wattal et al. \cite{wattal2012s} put products that the customer might like most in the email. By analyzing 30 email-marketing campaigns, Rettie \cite{rettie2002email} found that response rate is negatively correlated with email length. Carvalho \cite{carvalho2006personalization} proposed a personalization algorithm that put news liked by similar users in e-newsletters.

\item Order of contents: Besides the theories supporting putting important information on top \cite{shrestha2007eye}, there were also many studies supporting different arrangements. Wojdynski and Nathanie \cite{wojdynski2016going} examined 12 web page designs and found that the advertisements in the middle or bottom positions got better recognition. Heinz and Mekler \cite{heinz2012influence} found that banner placement did not influence recognition and recall.

\item Visual designs: Several studies focused on how to highlight the important content. Rettie \cite{rettie2002email} found that response rate is positively correlated with the number of images. Wojdynski and Nathanie \cite{wojdynski2016going} found that \textit{``users tend to gaze at a target object that is surrounded by objects with weaker “demand for attention” values.''}  

\end{enumerate}

\hlc[white]{We focused on leveraging personalization to improve bulk email's performance in this paper. But it is worth noting that there are other potential factors influencing recipients' engagement with bulk emails. For example, the email marketing platform Mailchimp found that sending frequency, writing (like the use of emojis), and sender's industry influence bulk email's open rate (ranging from 15\% to 28\% according to their report) \mbox{\cite{chimp2018average}}. Bulk email's from lines, signature lines \mbox{\cite{jenkins2008truth}}, and sending time and day \mbox{\cite{abrahams2010multi, bilovs2016open}} are also found to influence open rate by around 10\% in the previous experiments from marketing, communication, and management science.}

\subsection{Gaps}

\hlc[white]{Personalizing organizational bulk email is different from personal, commercial bulk email in several ways. Most important,
employees may have an obligation to read, know, and act upon information from their employer --- even information
they may personally not find interesting --- in a manner that does not apply to typical commercial bulk email. While
commercial bulk email may use one-off branding (to focus on one-time response rates) or recurring branding (to 
build a reputation and encourage repeat reading), organization bulk email nearly always uses recurring branding
tied to the structure and leadership of the organization.} Accordingly, organizations always have to balance the
tactical goal of having employees read the messages they choose to send (and view as important) while also
maintaining the effectiveness of the communication channel by having employees perceive the messages as 
relevant. There is no study on how different ways of personalization could help organizations achieve these two types of goals. In the following, we will introduce a field experiment we conducted to bridge this gap. 



\section{Background}
\subsection{Study Site and Newsletter}
The study site is a public university with over 25,000 employees and several campuses. There are communication offices in both the central units (e.g., presidential offices) and decentralized units (e.g., collegiate offices). There are newsletter editors in central units in charge of collecting news, designing, and distributing university-wide newsletters.

Before the study, we met with three communicators from the central offices and decided to experiment with the newsletter U of M Brief (see figure \ref{fig:brief}), which is sent to all the employees weekly (with subject line \textit{``U of M Brief <Date>''}). Each brief contained around 30 messages and 7 sections. Brief encourages people to submit information about ``need-to-know'' administrative news, making the university more accessible and creating connections among faculty and staff, promoting healthy lives, or the university’s mission of outreach, research, teaching, and education.

\subsection{Communication Goals}
\hlc[white]{Based on the studied site and newsletter above, we now frame the organization's communication goals with Brief. The tactical goals are about how well the specific group of messages are recognized and read. The strategic goals measure employees' overall experiences with Brief. Specifically, we measured these 6 metrics:}

\noindent$\star$ \hlc[white]{Recognition/read-in-detail rate: the percentage of the investigated messages being self-reported as seen/read-in-detail by the employees in the study's end survey (tactical goal). For example, the percentage of the messages in Top News the employees reported ``seen'' when Top News were all organization-preferred messages. \footnote{We could not measure the reading time of a single message in Brief because of the technical challenge of tracking specific regions' reading time naturally (our future work). First, many browsers (e.g., Chrome and Gmail) block access to the exact loading time of invisible pixels. Second, there is a lack of low-cost eye-tracking technology (e.g., eye-tracking based on a single computer camera) \cite{ferhat2016low}; and employees might also pay more attention to bulk emails when being recorded by camera.}}

\noindent$\star$ \hlc[white]{Open rate: the percentage of the investigated Briefs being opened by the employees (strategic goal). For example, the percentage of a Brief being opened when we put the organization-preferred messages on subject lines.}

\noindent$\star$ \hlc[white]{Interest rate: the percentage of the investigated Briefs being rated as ``interesting'' by the employees (strategic goal). }

\noindent$\star$ \hlc[white]{Reading time: the average reading time of the investigated Briefs (strategic goal). }

\noindent$\star$ \hlc[white]{Overall recognition rate: the average of the recognition rates of all the investigated Briefs' messages (strategic goal). }


\subsection{Designs}
\hlc[white]{We considered 5 kinds of personalization designs --- original/random/employee-preferred/organization-preferred/mixed designs. We will discuss how we selected the employee/organization-preferred messages in 4.2. Let us now use Figure \mbox{\ref{fig:brief}}'s Brief as an example. Suppose a faculty member is interested in biology stories and sports while the organization wants them to know about administrative and social justice updates.}

\noindent\textbf{A Subject line}: which message to be added to the Brief's subject line.

\textbf{A1} Original subject line: \textit{``U of M Brief (October 27, 2021)''}.

 \textbf{A2} Random subject line: the original subject line with a random message, e.g., \hlc[white]{\textit{``U of M Brief (October 27, 2021) - Fall 2021 Capstone presentations''}}.

 \textbf{A3} Organization-preferred subject line: the original subject line with the message that the organization mostly preferred the faculty member to read, e.g., \hlc[white]{\textit{``U of M Brief (October 27, 2021) - Board of Regents meeting highlights''}}.

 \textbf{A4} Employee-preferred subject line: the original subject line with the message that the faculty member mostly preferred, e.g., \hlc[white]{\textit{``U of M Brief (October 27, 2021) - Men's hockey return to ACHA''}}.

\noindent\textbf{B Top news}: which 4 messages are to be selected as the Brief's top news.

\textbf{B1} Original top news: the same top news as Figure \ref{fig:brief}.

\textbf{B2} Random top news: use 4 random messages.

\textbf{B3} Organization-preferred top news: use the 4 messages the organization most preferred the faculty member to know. \hlc[white]{E.g., Board of Regents meeting highlights, Work for social justice, etc}.

\textbf{B4} Employee-preferred top news: use the 4 messages the faculty member mostly preferred as top news. \hlc[white]{E.g., Men's hockey return to ACHA, A key biological pathway, etc}.

\textbf{B5} Mixed top news: mix 2 employee-preferred messages and 2 organization-preferred messages in top news. \hlc[white]{E.g., Men's hockey return to ACHA, Board of Regents meeting highlights, A key biological pathway, Work for social justice}.

\noindent\textbf{C Message Order}: how to sort the messages in the non-top sections of this Brief.

\textbf{C1} Original order: use the original Brief's messages' order.

\textbf{C2} Random order: sort the messages randomly.

\textbf{C3} Organization-preferred order: sort the messages by the organization's preference.

\textbf{C4} Employee-preferred order: sort the messages by the faculty member's preference.

\textbf{C5}\hlc[white]{ Zipper order: repeat this process --- select the message with the highest employee-preference score (see 4.2), the message with the highest organization-preference score, the message with the 2nd highest employee-preference, etc}.

If a message is added to the subject line but not selected to top news, we add it to the end of top news to avoid the employees feeling deceived if they click into the Briefs because of the subject lines. If a message from the campus sections was selected to top news, the name of its campus would be added to its title.

Within each treatment, we had two control groups: a good original control group (A1, B1, C1) which used the original subject lines/top news/message order --- as we discussed, these were carefully selected by an experienced editor (the communicator of Brief) according to their criterion on how to design Briefs; a bad random control group (A2, B2, C2) which used random subject lines/top news/message order generated by the system. Figure \ref{fig:per_brief} is a sample personalized Brief for this faculty member if we assigned them to A4 x B5 x C5. 


\subsection{Research Questions and Hypotheses}

We proposed hypotheses and questions on these designs' influences on the communication goals. 

\noindent\textbf{A Subject lines.} When subject lines match the employees' preferences, Brief might achieve these strategic goals: 

\noindent  Adding employee-preferred messages on subject lines will increase the newsletter's interest rate (\textbf{H1.1}) / reading time (\textbf{H1.2}) / overall recognition rate (\textbf{H1.3}) / open rate  (\textbf{H1.4}).

On tactical goals, the messages on subject lines should have a greater chance of being seen by the employees:

\noindent  (\textbf{H1.5}) Putting messages on subject lines will increase these messages' recognition rates.

We expect only to see an improvement in the read-in-detail rate for those employee-preferred messages, as the content should be interesting to employees to make them click/read in detail \cite{kim2016click, kessler2019we}:

\noindent  (\textbf{H1.6}) Putting employee-preferred messages on subject lines will increase their read-in-detail rates. 

\noindent\textbf{B Top News.} When top news matched the employees' preferences, Brief might achieve a better interest rate:

\noindent \textbf{H2.1} Putting employee-preferred messages in top news will increase the newsletter's interest rate.

We do not have theories to predict the effect of placing employee-preferred messages in top news on the reading time and overall recognition rate. For example, when putting employee-preferred messages in top news, employees might be motivated to read the rest of the newsletter or only read top news and leave. We proposed these questions:

\noindent What is the effect of putting organization-preferred messages/employee-preferred messages/mixing employee-preferred messages and organization-preferred messages in top news on the newsletter's reading time (\textbf{Q2.2}) and overall recognition rate (\textbf{Q2.3})?


We hope that the mixed top news let employees read the organization-preferred messages when reading interesting messages: Mixing organization-preferred messages with the employee-preferred messages in top news will increase the organization-preferred messages' recognition rates (\textbf{H2.4}).

Besides that, we also had the hypotheses similar to those for the subject lines on tactical goals:

\noindent  (\textbf{H2.5}) Putting messages in top news will increase their recognition rates.

\noindent  (\textbf{H2.6}) Putting employee-preferred messages in top news will increase their read-in-detail rates. 

\noindent\textbf{C Message Order.} We were uncertain about the direction and scale of message order's effect. We proposed the following questions: what is the effect of sorting messages by employee's preference/organization's preference/zipper order on the overall recognition rate (\textbf{Q3.1}) / reading time (\textbf{Q3.2}) of the newsletter?

\noindent(\textbf{Q3.3}) What is the effect of interleaving messages (sorting messages by the zipper order of employee/organization's preference) on the recognition rates of the organization-preferred messages?

We hope that sorting by employee's preference would make them feel the Brief is more interesting:

\noindent(\textbf{H3.1}) Sorting messages by employee's preference will increase the interest rate of the newsletter.

We summarized our hypotheses and research questions in Table \ref{tab:rq}. There were blanks (marked as to be observed) in this table when we did not find any related work or reasons to make a guess on a significant effect. We just observed what happened in these blanks. Besides the questions above.

\begin{table}[!htbp]
\caption{\hlc[white]{Summary of the newsletter's communication goals, hypotheses, and research questions. org-pref messages: organization-preferred messages. emp-pref messages: employee-preferred messages. A1/2, B1/2, C1/2 are control groups.}}~\label{tab:rq}
\centering
\resizebox{\textwidth}{!}{%
\begin{tabular}{|l|l|l|l|l|l|l|l|} 
\hline
                                                                              &                                                                                 & \multicolumn{4}{c|}{\textbf{Strategic Goal }}                                                                                                                                                                                                                                                                                         & \multicolumn{2}{c|}{\textbf{Tactical Goal }}                                                                                                                                                                                           \\ 
\hline
\textbf{Group}                                                                & \textbf{Treatment}                                                              & \begin{tabular}[c]{@{}l@{}}Interest Rate \\of this Brief\end{tabular}           & \begin{tabular}[c]{@{}l@{}}Reading time \\of this Brief\end{tabular}              & \begin{tabular}[c]{@{}l@{}}Recognition Rate \\of this Brief\end{tabular}                      & \begin{tabular}[c]{@{}l@{}}Open Rate \\of this Brief\end{tabular} & \begin{tabular}[c]{@{}l@{}}Recognition Rate \\of this message\end{tabular}                                                                       & \begin{tabular}[c]{@{}l@{}}Read-in-detail Rate \\of this message\end{tabular}       \\ 
\hline
\multirow{4}{*}{\begin{tabular}[c]{@{}l@{}}A: \\Subject \\lines\end{tabular}} & 1: Original                                                                     & \begin{tabular}[c]{@{}l@{}}To be\\observed\end{tabular}                         & \begin{tabular}[c]{@{}l@{}}To be\\observed\end{tabular}                           & \begin{tabular}[c]{@{}l@{}}To be\\observed\end{tabular}                                       & \begin{tabular}[c]{@{}l@{}}To be\\observed\end{tabular}           & \begin{tabular}[c]{@{}l@{}}To be\\observed\end{tabular}                                                                                          & \begin{tabular}[c]{@{}l@{}}To be\\observed\end{tabular}                             \\ 
\cline{2-8}
                                                                              & \begin{tabular}[c]{@{}l@{}}2: Add a \\random\\message\end{tabular}              & \begin{tabular}[c]{@{}l@{}}To be\\observed\end{tabular}                         & \begin{tabular}[c]{@{}l@{}}To be~\\observed\end{tabular}                          & \begin{tabular}[c]{@{}l@{}}To be\\observed\end{tabular}                                       & \begin{tabular}[c]{@{}l@{}}To be\\observed\end{tabular}           & \begin{tabular}[c]{@{}l@{}}To be\\observed\end{tabular}                                                                                          & \begin{tabular}[c]{@{}l@{}}To be\\observed\end{tabular}                             \\ 
\cline{2-8}
                                                                              & \begin{tabular}[c]{@{}l@{}}3: Add an \\org-pref \\message\end{tabular}          & \begin{tabular}[c]{@{}l@{}}To be \\observed\end{tabular}                        & \begin{tabular}[c]{@{}l@{}}To be \\observed\end{tabular}                          & \begin{tabular}[c]{@{}l@{}}To be \\observed\end{tabular}                                      & \begin{tabular}[c]{@{}l@{}}To be \\observed\end{tabular}          & \begin{tabular}[c]{@{}l@{}}H1.5 Increase  \\recognition rate.\end{tabular}                                                                  & \begin{tabular}[c]{@{}l@{}}To be \\observed\end{tabular}                            \\ 
\cline{2-8}
                                                                              & \begin{tabular}[c]{@{}l@{}}4: Add an \\emp-pref \\message\end{tabular}          & \begin{tabular}[c]{@{}l@{}}H1.1 Increase \\ interest \\rate.\end{tabular} & \begin{tabular}[c]{@{}l@{}}H1.2 Increase \\reading time.\end{tabular}             & \begin{tabular}[c]{@{}l@{}}H1.3 Increase \\overall \\recognition rate.\end{tabular}            & \begin{tabular}[c]{@{}l@{}}H1.4 Increase\\open rate.\end{tabular} & \begin{tabular}[c]{@{}l@{}}H1.5 Increase  \\recognition rate.\end{tabular}                                                                  & \begin{tabular}[c]{@{}l@{}}H1.6 Increase  \\read-in-detail rate.\end{tabular}  \\ 
\hline
\multirow{5}{*}{\begin{tabular}[c]{@{}l@{}}B: \\Top \\news\end{tabular}}      & \begin{tabular}[c]{@{}l@{}}1: Original\\top news\end{tabular}                   & \begin{tabular}[c]{@{}l@{}}To be\\observed\end{tabular}                         & \begin{tabular}[c]{@{}l@{}}To be\\observed\end{tabular}                           & \begin{tabular}[c]{@{}l@{}}To be\\observed\end{tabular}                                       &                                                                   & \begin{tabular}[c]{@{}l@{}}To be\\observed\end{tabular}                                                                                          & \begin{tabular}[c]{@{}l@{}}To be\\observed\end{tabular}                             \\ 
\cline{2-8}
                                                                              & \begin{tabular}[c]{@{}l@{}}2: Put \\random\\messages\end{tabular}               & \begin{tabular}[c]{@{}l@{}}To be\\observed\end{tabular}                         & \begin{tabular}[c]{@{}l@{}}To be\\observed\end{tabular}                           & \begin{tabular}[c]{@{}l@{}}To be\\observed\end{tabular}                                       &                                                                   & \begin{tabular}[c]{@{}l@{}}To be\\observed\end{tabular}                                                                                          & \begin{tabular}[c]{@{}l@{}}To be\\observed\end{tabular}                             \\ 
\cline{2-8}
                                                                              & \begin{tabular}[c]{@{}l@{}}3: Put \\org-pref \\messages\end{tabular}            & \begin{tabular}[c]{@{}l@{}}To be \\observed\end{tabular}                        & \begin{tabular}[c]{@{}l@{}}Q2.2 How does\\it affect\\reading time?\end{tabular} & \begin{tabular}[c]{@{}l@{}}Q2.3 How does it\\affect overall \\recognition rate?\end{tabular} &                                                                   & \begin{tabular}[c]{@{}l@{}}H2.5 Increase  \\recognition rate.\end{tabular}                                                                  & \begin{tabular}[c]{@{}l@{}}To be \\observed\end{tabular}                            \\ 
\cline{2-8}
                                                                              & \begin{tabular}[c]{@{}l@{}}4: Put \\emp-pref \\messages\end{tabular}            & \begin{tabular}[c]{@{}l@{}}H2.1 Increase \\ interest \\rate.\end{tabular} & \begin{tabular}[c]{@{}l@{}}Q2.2 How does\\it affect\\reading time?\end{tabular} & \begin{tabular}[c]{@{}l@{}}Q2.3 How does\\it affect  overall\\recognition rate?\end{tabular} &                                                                   & \begin{tabular}[c]{@{}l@{}}H2.5 Increase  \\recognition rate.\end{tabular}                                                                  & \begin{tabular}[c]{@{}l@{}}H2.6 Increase  \\read-in-detail rate.\end{tabular}  \\ 
\cline{2-8}
                                                                              & \begin{tabular}[c]{@{}l@{}}5: Mix \\emp-pref\\/org-pref \\messages\end{tabular} & \begin{tabular}[c]{@{}l@{}}To be \\observed\end{tabular}                        & \begin{tabular}[c]{@{}l@{}}Q2.2 How does\\it affect\\reading time?\end{tabular} & \begin{tabular}[c]{@{}l@{}}Q2.3 How does it\\affect overall \\recognition rate?\end{tabular} &                                                                   & \begin{tabular}[c]{@{}l@{}}H2.4 Increase  \\recognition rate \\of org-pref messages. \\H2.5 Increase  \\recognition rate.\end{tabular} & \begin{tabular}[c]{@{}l@{}}To be \\observed\end{tabular}                            \\ 
\hline
\multirow{5}{*}{\begin{tabular}[c]{@{}l@{}}C: \\Order\end{tabular}}           & \begin{tabular}[c]{@{}l@{}}1: Original\\order\end{tabular}                      & \begin{tabular}[c]{@{}l@{}}To be\\observed\end{tabular}                         & \begin{tabular}[c]{@{}l@{}}To be\\observed\end{tabular}                           & \begin{tabular}[c]{@{}l@{}}To be\\observed\end{tabular}                                       &                                                                   & \begin{tabular}[c]{@{}l@{}}To be\\observed\end{tabular}                                                                                          & \begin{tabular}[c]{@{}l@{}}To be\\observed\end{tabular}                             \\ 
\cline{2-8}
                                                                              & \begin{tabular}[c]{@{}l@{}}2: Random\\order\end{tabular}                        & \begin{tabular}[c]{@{}l@{}}To be\\observed\end{tabular}                         & \begin{tabular}[c]{@{}l@{}}To be\\observed\end{tabular}                           & \begin{tabular}[c]{@{}l@{}}To be\\observed\end{tabular}                                       &                                                                   & \begin{tabular}[c]{@{}l@{}}To be\\observed\end{tabular}                                                                                          & \begin{tabular}[c]{@{}l@{}}To be\\observed\end{tabular}                             \\ 
\cline{2-8}
                                                                              & \begin{tabular}[c]{@{}l@{}}3: Sort \\by org\\-preference\end{tabular}           & \begin{tabular}[c]{@{}l@{}}To be \\observed\end{tabular}                        & \begin{tabular}[c]{@{}l@{}}Q3.1 How does\\it affect \\reading time?\end{tabular}  & \begin{tabular}[c]{@{}l@{}}Q3.2 How does it\\affect overall \\recognition rate?\end{tabular}   &                                                                   &                                                                                                                                                  &                                                                                     \\ 
\cline{2-8}
                                                                              & \begin{tabular}[c]{@{}l@{}}4: Sort \\by emp\\-preference\end{tabular}           & \begin{tabular}[c]{@{}l@{}}H3.1 Increase \\ interest \\rate.\end{tabular} & \begin{tabular}[c]{@{}l@{}}Q3.1 How does\\it affect \\reading time?\end{tabular}  & \begin{tabular}[c]{@{}l@{}}Q3.2 How does it\\affect overall\\recognition rate?\end{tabular}   &                                                                   &                                                                                                                                                  &                                                                                     \\ 
\cline{2-8}
                                                                              & \begin{tabular}[c]{@{}l@{}}5: Sort by \\mix-order\end{tabular}                  & \begin{tabular}[c]{@{}l@{}}To be \\observed\end{tabular}                        & \begin{tabular}[c]{@{}l@{}}Q3.1 How does\\it affect \\reading time?\end{tabular}  & \begin{tabular}[c]{@{}l@{}}Q3.2 How does it\\affect overall\\recognition rate?\end{tabular}   &                                                                   & \begin{tabular}[c]{@{}l@{}}Q3.3 How does it\\affect  recognition rate\\of org-pref messages?\end{tabular}                                &                                                                                     \\
\hline
\end{tabular}
}
\end{table}

\section{Methods}
We collaborated with the editor of Brief to conduct the experiment in the university with 141 employees. The experiment took 8 weeks. The steps were (see Figure \ref{fig:procedure}):

\noindent \textbf{Step 1 (week 1), recruiting and assigning participants}: through Brief and a communication newsletter. \hlc[white]{Each selected participant would be assigned to a treatment combination $A_i\times B_j\times C_k$ through the whole study}.

\noindent \textbf{Step 2 (week 1), collecting and calculating the employees' preferences}: \hlc[white]{the employees filled in preference surveys and we collected their work-relevance/interest scores for each topic}.

\noindent \textbf{Step 3 (week 2 to 7), collecting and calculating the organization's preferences}:  \hlc[white]{the editor sent the draft of the newsletter to the experiment system every week. The system extracted text and html, then sent the editor a survey to collect each message's topics (up to 4) and work-relevance/importance scores from the organization's perspective}.

\noindent \textbf{Step 4 (week 2 to 7), generating newsletters}: the system sent the original non-personalized newsletter to collect base performance data in week 2 and generated personalized newsletters based on the employees' experimental groups, organization's preference, and employee's preference during weeks 3 to 7. 

\noindent \textbf{Step 5 (week 8), collecting performance metrics and feedback}: the system sent out the end surveys to collect the recognition/read-in-detail/interest data, and the log files of a plugin which tracked open rates and reading time.




\begin{figure}[!htbp]
\centering
  \includegraphics[width=0.8\columnwidth]{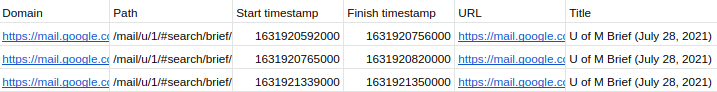}
  \caption{A sample csv file recorded by the plugin. It records the domain, path (which helps us to discriminate whether it is the experimental Brief), start/end reading timestamp, url, and tab titles.}~\label{fig:plugin}
\end{figure}

\subsection{Recruitment (step 1, week 1)}
\hlc[white]{The goal of this experiment is to study newsletter reading in a natural context. The scope of this experiment is organizations' newsletters sent to a large number of employees and their regular and occasional readers. We did not study employees who never read Briefs because reading the Brief is not their natural behavior --- being in the experiment could force their open rate and recognition rate to increase from 0 to a substantial level, which could largely interfere with the main outcomes of our interventions. Accordingly, we posted the recruitment message as the first message of a Brief. In the hope of broadening our participants, we also reached out to the communicator network and posted the message at the top of its newsletter \footnote{We did two things to understand the generalizability in view of the Brief-centric recruitment --- 1) compare the participants from different sources: the 10 other employees recruited from the communication newsletter had a base recognition rate of the week 2's original Brief (29\%) similar to the other 107 employees recruited from Brief (30\%); 2) checking the number of occasional Brief readers: 44 of our participants did not open all the experimental Briefs, and 22 participants opened less than or equal to 60\% of the experimental Briefs (Brief's average open rate).}. The newsletters only had their brands without specific messages on their subject lines that could be used to target specific audiences \footnote{The messages closest to our recruitment message in that recruitment Brief had topics on Administrative News/Student Stories/Awards and Recognition. According to Table \mbox{\ref{tab:conflict}}, our participants were not more interested in these topics compared to other topics.}. }

\hlc[white]{We planned to have at least 100 participants. This number was estimated through 1000 simulations on the generalized linear mixed model power analysis tool SMIR \mbox{\cite{green2016simr}}. The analysis of open rate/interest rate, where we could only collect 5 data points from each employee, has the highest requirement. We target to observe a 15\% change in these metrics with a 20\% standard deviation and an 80\% power. Considering the dropout rate, we targeted 140 to 150 participants.}

In the signup form, we asked the potential participants whether they were employees of the university and mainly used Gmail and Chrome in reading Briefs and selected those confirmed. We also got their campuses and job categories. To make our surveys more concise, 20 job families of the university's human resource system were summarized into 7 job categories by two researchers and the Brief editor, according to whether these categories were considered different audience groups of Brief (see Appendix B).

We received 304 responses to our recruitment message and selected participants from this pool. We balanced the number of selected participants from different campuses, job categories, and recruiting sources and contacted 181 employees to set up the study (\hlc[white]{38 did not reply and were not enrolled}). Each employee filled in a preference survey on message's topics and had a 20-minute 1-to-1 zoom meeting with a research team member. In this meeting, we helped the employee: 1) set up a filter rule in their university Gmail, which archived all the original Briefs they received during the experiment into a separate folder. They were told to avoid checking this folder during the study; 2) install a plugin on their Chrome browser. The plugin only recorded the time they spent when they were on a tab with the text ``U of M Brief'' (see figure \ref{fig:plugin}). 141 participants completed the setup process (2 employees could not install the plugin and were not enrolled). The participants were compensated with a \$20 Amazon gift card after setting up the study.
\begin{table}[!htbp]
\centering
\caption{\hlc[white]{The variables used in the personalization procedure and their definitions. Except that $org\_importance_{k}$ is the same for all the employees, the employee/organization's preferences were calculated by the inputs from the employees/editor on the left column.}}
\label{tab:symbol}
\arrayrulecolor[rgb]{0.753,0.753,0.753}
\scalebox{0.70}{
\begin{tabular}{!{\color{black}\vrule}l|l!{\color{black}\vrule}l|l!{\color{black}\vrule}} 
\arrayrulecolor{black}\hline
\begin{tabular}[c]{@{}l@{}}\textbf{Inputs from }\\\textbf{Employees}\end{tabular}                & \textbf{\textbf{Definition}}                                                                                                          & \begin{tabular}[c]{@{}l@{}}\textbf{Employee's }\\\textbf{Preference}\end{tabular}                                                           & \textbf{Definition}                                                                                                                                                   \\ 
\arrayrulecolor[rgb]{0.753,0.753,0.753}\hline
$campus_i$                                                                                       & $employee_i$'s campus                                                                                                                 & $emp\_interest_{i,k}$                                                                                                                       & \begin{tabular}[c]{@{}l@{}}$employee_i$'s interest score for~$message_k$~(0 to 1)\end{tabular}                                                                      \\ 
\hline
$job_i$                                                                                          & $employee_i$'s job category   (Appendix B)                                                                                                        & $emp\_relevance_{i,k}$                                                                                                                      & \begin{tabular}[c]{@{}l@{}}$employee_i$'s work-relevance score for $message_k $\\(0 to 1)\end{tabular}                                                                \\ 
\hline
$interest_{i,j}$                                                                                 & \begin{tabular}[c]{@{}l@{}}whether~$employee_i$~looks for interesting \\messages from $topic_j$ (0 or 1)\end{tabular}                                    & $emp\_pref_{i,k}$                                                                                                                           & \begin{tabular}[c]{@{}l@{}}$employee_i'$s preference on $message_k$ \\($=(emp\_interest_{i,k}+emp\_relevance_{i,k})/2$, 0 to 1)\\\end{tabular}                           \\ 
\hline
$relevance_{i,j}$                                                                                & \begin{tabular}[c]{@{}l@{}}whether~$employee_i$~looks for work-relevant \\messages from~$topic_j$~(0 or 1)\end{tabular}               &                                                                                                                                             &                                                                                                                                                                       \\ 
\arrayrulecolor{black}\hline
\begin{tabular}[c]{@{}l@{}}\textbf{\textbf{Inputs from }}\\\textbf{\textbf{Editor}}\end{tabular} & \textbf{\textbf{Definition}}                                                                                                          & \begin{tabular}[c]{@{}l@{}}\textbf{\textbf{\textbf{\textbf{Organization's }}}}\\\textbf{\textbf{\textbf{\textbf{Preference}}}}\end{tabular} & \textbf{\textbf{\textbf{\textbf{Definition}}}}                                                                                                                        \\ 
\arrayrulecolor[rgb]{0.753,0.753,0.753}\hline
$important_k$                                                                                    & \begin{tabular}[c]{@{}l@{}}$message_k'$s general importance to all the\\employees from the organization's view\\(1 to 4)\end{tabular} & $org\_importance_{k}$                                                                                                                       & \begin{tabular}[c]{@{}l@{}}$message_k$'s general importance score to all\\the employees~ from the organization's view\\($=(important_k-1)/3$, 0 to 1)\end{tabular}  \\ 
\hline
$target\_campus_k$                                                                               & $message_k$'s target campuses (list)                                                                                                  & $org\_relevance_{i,k}$                                                                                                                      & \begin{tabular}[c]{@{}l@{}}organization's work-relevance score for \\$message_k$ given~$employee_i$~(0 to 1)\end{tabular}                                             \\ 
\hline
$target\_job_k$                                                                                  & $message_k$'s target job categories~(list)                                                                                            & $org\_pref_{i,k}$                                                                                                                           & \begin{tabular}[c]{@{}l@{}}organization's preference on~$message_k$ given\\ $ employee_i$\\($=(org\_relevance_{i,k} + org\_importance_{k})/2$, 0 to 1)\end{tabular}                                                        \\ 
\hline
\begin{tabular}[c]{@{}l@{}}$topic\_list_k$\\\end{tabular}                                        & \begin{tabular}[c]{@{}l@{}}$message_k$'s relevant topics (list, up to 4,\\ Appendix A)\end{tabular}                                                                                         &                                                                                                                                             &                                                                                                                                                                       \\
\arrayrulecolor{black}\hline
\end{tabular}
}
\end{table}

\subsection{Personalization Procedure}
To define and collect preferences data within limited survey questions, we used ``topic'' as a bridge to connect messages
and preferences (we assume that each message could have at most 4 topics). 20 topics were summarized by a thematic analysis \cite{braun2012thematic} of 140 messages from 5 previous Briefs (see Appendix A). 5 research team members grouped them, labeled the clusters, and identified the hierarchy. The Brief editor checked the list and suggested two special topics --- ``news from my campus'' and ``news from other campuses''. We summarized the symbols we used in the personalization procedure and their definitions in Table \ref{tab:symbol}.



\subsubsection{\hlc[white]{Collecting and calculating employees' preferences (step 2, week 1)}}
At the set up of this study, we asked the participants to fill out a preference survey (figure \ref{fig:procedure} - week 1). For each topic, we asked the participants to check whether the statements ``I would look up this category for messages interesting to me'' or ``I would look up this category for messages work-relevant to me'' applied to them separately. Let's call $employee_i$'s answers to $topic_j$ as $interest_{i,j}$ and $relevance_{i,j}$, which took the value 0 or 1. We also asked the campus ($campus_i$) and job category ($job_i$) in the survey.

\hlc[white]{Then for $employee_i$ and $message_k$, we calculated the employee's preference on this message ($emp\_pref_{i,k}$) with 3  steps. First, $employee_i$'s interest score for $message_k$ is defined as (see 4.2.2 for $topic\_list_k$ and $target\_campus_k$)}:
\begin{small}
\begin{equation}
    emp\_interest_{i,k} =
    \begin{dcases}
    0, \text{if } employee_i \ \text{is not interested in other campuses' messages and } campus_i \\ \quad \centernot\in target\_campus_k \\ 
     avg(\{interest_{i,q}|topic_q \in topic\_list_k\}),               \text{otherwise}
\end{dcases}
\end{equation}
\end{small}

\hlc[white]{Second, we calculated $employee_i$'s work-relevance score $emp\_relevance_{i,k}$ for $message_k$ as}:
\begin{small}
\begin{equation}
    emp\_relevance_{i,k} =
    \begin{dcases}
    0, \text{if } employee_i \ \text{does not look for other campuses' relevant messages or  }\\ \quad campus_i \centernot\in target\_campus_k  \\
     avg(\{relevance_{i,q}|topic_q \in topic\_list_k\}),               \text{otherwise}
\end{dcases}
\end{equation}
\end{small}

\hlc[white]{Third,  $employee_i$'s preference on $message_k$ is defined as }
\begin{small}
\begin{equation}
    emp\_pref_{i,k} =  (emp\_interest_{i,k} + emp\_relevance_{i,k})/2
\end{equation}
\end{small}

\subsubsection{\hlc[white]{Collecting and calculating the organization's preferences (step 3, week 2 to 7)}}
During weeks 2 to 7, the Brief editor provided the organization's preference with each message weekly (Figure \ref{fig:procedure}, week 2 to 7 (1) - (5)). The Brief editor sent the draft Brief to the system every week. The system then retrieved the messages (subject lines, titles, content, html, etc.) from the draft Brief, generated the editor survey, and sent it to the editor. The system listened to the editor's responses and loaded the responses to the database. After the responses were successfully loaded, a verification message would be sent to the editor to ensure that the original Briefs could only be sent after the system got the data to generate personalized Briefs. The editor survey contained the following questions for each $message_k$:
\begin{enumerate}
    \item How relevant this message is in building community, pride, common understandings of excellence and mission of the university (from 1: not relevant to 4: very relevant)? ($importance_k$).  
    \item Select the employee categories that might find the message above work-relevant ($target\_job_k$).
    \item Specify the message's relevant topics (select no more than 4 topics) ($topic\_list_k$).
    \item There is an implicit question on target campus ($target\_campus_k$), as the editor suggested that the original Briefs' campus sections had already represented it.
\end{enumerate}

\hlc[white]{Then we calculate the organization's preference ($org\_pref_{i,k}$) on $message_k$ given $employee_i$ by 3 steps. First, the organization's work-relevance score for $message_k$ given $employee_i$ with job category ($job_i$) and campus ($campus_i$) is}:
\begin{small}
\begin{equation}
    org\_relevance_{i,k} = 1,              if \ campus_i \in target\_campus_k \ \text{and} \ job_i \in target\_job_k; \ 0, otherwise
\end{equation}
\end{small}

\hlc[white]{Second, $message_k$'s general importance score to all the employees is just the standardization of $important_k$. This value is the same for all the employees, as the editor suggested that the important messages should apply to all}:
\begin{small}
\begin{equation}
    org\_importance_{k} = (important_k-1)/3
\end{equation}
\end{small}

\hlc[white]{Third, the organization's preference on $message_k$ given $employee_i$ is defined as}:
\begin{small}
\begin{equation}
    org\_pref_{i,k} =  (org\_relevance_{i,k} + org\_importance_{k})/2
\end{equation}
\end{small}


\subsubsection{Generating newsletters (step 4, week 2 to 7)}
After we calculated these preference data, the system generated personalized Briefs from weeks 3 to 7 (original Brief for week 2) and sent these Briefs to the employees. We give the employees the choice to select to receive the email at 6 AM or 9 AM, given which time is better for them to receive Briefs and read it on their laptop or desktop's Chrome with the plugin we installed (eventually, we did not observe significant differences on the performance metrics between the 6 AM and 9 AM group).

\begin{figure}[!htbp]
\centering
  \includegraphics[width=0.9\columnwidth]{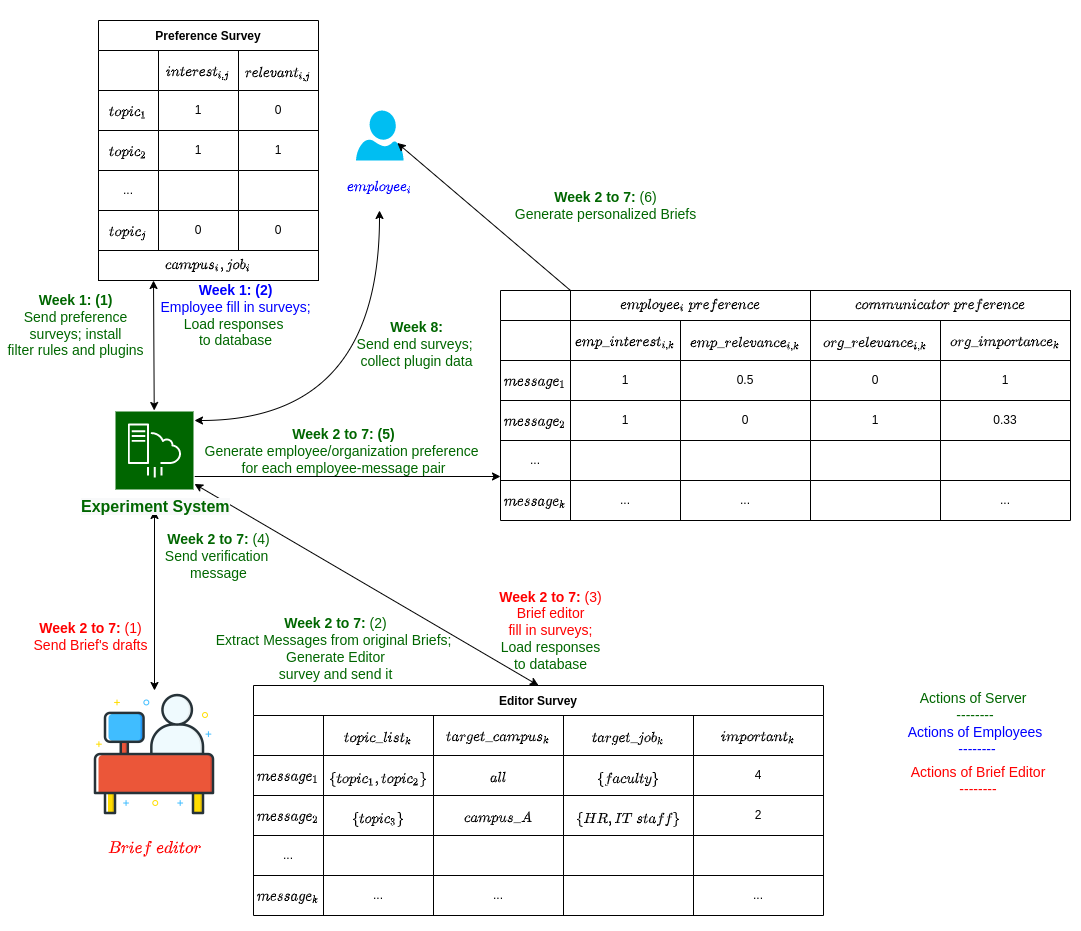}
  \caption{The outline of the experiment procedure. Weeks 2 to 7's steps were repeated weekly. We sent the original Brief in week 2.}~\label{fig:procedure}
\end{figure}

\hlc[white]{With a 4 x 5 x 5 factorial design on the treatments A (subject lines), B (top news), C (message order) below, each participant would be assigned to a treatment combination $A_i \times B_j \times C_k$ through the study randomly. Their Briefs' would be generated according to the criterion in 4.2}, based on the employee preference $emp\_pref$ and organization preference $org\_pref$ we calculated for each message (Figure \ref{fig:procedure}, week 2 to 7 (6)).

\subsection{Collecting Performance Metrics and Feedback (step 5, week 8)}
At week 8, we sent out the end survey (Figure \ref{fig:procedure}, week 8). Participants were compensated with another \$20 Amazon gift card after submitting the end surveys. We collected the recognition data for the message below: 1) week 2 and week 7's messages in the top news, up to the top 10 messages in the u-wide news, up to the top 2 messages in the participant's campus news; 2) weeks 5 to 7's messages in the top news; 3) weeks 3 to 7's messages on the subject lines.

The recognition data was collected by the question ``Have you seen it in recent Briefs? No/Not Sure/Skimmed/Read fully''. We defined $employee_i$'s \textbf{$recognition$} as 1 if the answer is skimmed or read fully, and \textbf{$read\_in\_detail$} as 1 if the answer is read fully. After that, the survey asked the participants to indicate how interesting each Brief is to them in general from ``1 Not interesting'' to ``4 Very interesting''. The survey also collected plugin data. We would know the reading time of each Brief, and whether the participants opened a Brief or not. We also collected the interest scores (scale 1 to 4) and work-relevant scores (scale 1 to 4) for week 2's messages to study the size of conflicts. The order of these questions is randomized. The participants were asked not to search their inbox while answering these questions. For each experimental group (e.g., the Briefs with random subject lines), we reported and tested:

\noindent\hlc[white]{
 $\star$ \textbf{$recognition\_rate$} and \textbf{$read\_in\_detail\_rate$}: the percentage of the considered messages (e.g, the messages in the top news, the message on subject lines) in the experimental group that got $recognition = 1$ or $read\_in\_detail=1$.}

\noindent\hlc[white]{
 $\star$ \textbf{$overall\_recognition\_rate$}: the percentage of the messages in this experimental group's Briefs with $recognition = 1$.}

\noindent \hlc[white]{
 $\star$ \textbf{$interest\_rate$}: the percentage of Briefs that got an interesting level $\geq$ 3 in that experimental group.}

\noindent \hlc[white]{
$\star$ \textbf{$open\_rate$}: the percentage of Briefs that got $open = 1$ in that experimental group.}

\noindent\hlc[white]{
 $\star$ \textbf{$reading\_time$}: the average of the Briefs' reading time in that experimental group.}

We received 132 responses for the end survey, and 117 of them were complete. There were 15 incomplete responses either because the participants did not complete the surveys, the plugin was deleted or blocked by a Chrome update, or the participants lost access to their devices. This dataset contained the recognition data and read-in-detail data of 4242 messages, and the recognition data, open data, and reading time data of 702 Briefs in total. \hlc[white]{We did received 2 reports of participants forgetting to read in the browser, and their data was excluded.}

\hlc[white]{To avoid spillover effects \mbox{\cite{sinclair2012detecting}}, our participants were scheduled for separate 1-on-1 meetings in the setup, and they were not aware of each other's participation nor their experimental groups. We observed no communication or sharing that would have led to spillover effects. To avoid Hawthorne effects \mbox{\cite{jones1992there}}, we sent out the original Briefs in week 2, and measured the base performance data, which would be later included in our models as a factor. And our participants were in the experiment for 6 weeks, to avoid incentivizing participants to pay more attention to read/remember these Briefs, we only analyze the performance data collected in the last week of this experiment.}


\section{Results}
\subsection{Analysis and Overview}
We summarized the performance of each experimental group in Table \ref{tab:result_original} and the results on the hypotheses and questions in Table \ref{tab:result}. We built mixed logistic models to evaluate categorical performance metrics (interest rate, recognition rate, open rate, and read-in-detail rate) and mixed linear models to test numerical performance metrics (reading time) by the afex package, which provided ANOVA table with likelihood-ratio tests for both linear and logistic models \cite{singmann2015package}. We had a random effect based on subjects (from which employee we collected this data point) \cite{barr}, and we selected likelihood-ratio tests because we had many levels on the random effect (number of participants) \cite{barr2013random}. The independent variables include the corresponding experimental groups and the base performance metrics (the average of that performance metric given the corresponding employee's reactions with week 2's original Brief). For the base open rate, we used the number of Briefs they opened in 2021/the number of Briefs they received in 2021 before the experiment. We asked the employees to input queries in their Gmail in the preference survey to retrieve this number. If they have deleted Briefs, they reported their approximate numbers. We excluded the employees who gave the same  interest scores to all the experimental Briefs from the analysis of interest scores and excluded the employees who opened all or did not open any of the experimental Briefs from the analysis of open rates.

\begin{table}[!htbp]
\caption{\textbf{The performance metrics of each experimental group}, and the standard deviations of that group's participants' performance metrics. The blank cells are not applicable or not of interest. The employees who gave the same interest scores to all the experimental Briefs were excluded from the analysis of interest rates; those who opened all or did not open any of the experimental Briefs were excluded from the analysis of open rates. non-s: messages not on subject lines. non-t: messages not in top news.}~\label{tab:result_original}
\centering
\arrayrulecolor{black}
\scalebox{0.72}{
\begin{tabular}{|l|l|l|l|l|l!{\vrule width \heavyrulewidth}l|l|} 
\hline
                                  &                                                                                & \multicolumn{4}{c!{\vrule width \heavyrulewidth}}{\textbf{Strategic Goal}}                                                                                                                                                                                                                                                                                                                                                                                                                                                                              & \multicolumn{2}{c|}{\textbf{Tactical Goal}}                                                                                                                                                                                                                                                                                                \\ 
\hline
\begin{tabular}[c]{@{}l@{}}\textbf{ Group }\end{tabular}                  & \textbf{Treatment}                                                             & \begin{tabular}[c]{@{}l@{}}Interest Rate \\of Brief (\%) \end{tabular}                                                               & \begin{tabular}[c]{@{}l@{}}Reading Time \\of Brief \\(seconds)\end{tabular}                                                          & \begin{tabular}[c]{@{}l@{}}Recognition Rate \\of Brief (\%)\end{tabular}                                                                & \begin{tabular}[c]{@{}l@{}}Open Rate of\\Brief (\%)\end{tabular}                                          & \begin{tabular}[c]{@{}l@{}}Recognition Rate \\of Message (\%)\end{tabular}                                                                                                                   & \begin{tabular}[c]{@{}l@{}}Read-in-detail \\Rate of \\Message (\%)\end{tabular}                                                \\ 
\toprule
{ \begin{tabular}[c]{@{}l@{}}Non-s\\messages\end{tabular}} 
&\begin{tabular}[c]{@{}l@{}}\ \\ \ \\ \ \end{tabular} 
&&&&& \textbf{41\%}, 20\%&\textbf{15\%}, 13\% \\
\hline
\multirow{4}{*}{ \begin{tabular}[c]{@{}l@{}}A:\\Subject\\lines\end{tabular}} 
&\begin{tabular}[c]{@{}l@{}}1: Original \\subject \\line\end{tabular}         
& \begin{tabular}[c]{@{}l@{}}\textbf{68\%}, 22\% \end{tabular}   &  \textbf{143s}, 92s &\textbf{44\%}, 25\% &\begin{tabular}[c]{@{}l@{}}\textbf{65\%}, 21\%\end{tabular} & & \\\cline{2-8}
&\begin{tabular}[c]{@{}l@{}}2: Add a \\random \\message\end{tabular}         
&\begin{tabular}[c]{@{}l@{}}\textbf{69\%}, 27\% \end{tabular} & \textbf{150s}, 105s &\textbf{39\%}, 20\% &\begin{tabular}[c]{@{}l@{}}\textbf{63\%}, 14\%\end{tabular} &\textbf{40\%}, 27\% &\textbf{15\%}, 17\% \\\cline{2-8}
&\begin{tabular}[c]{@{}l@{}}3: Add an \\org-pref \\message\end{tabular}         
& \begin{tabular}[c]{@{}l@{}}\textbf{67\%}, 29\%  \end{tabular}& \textbf{149s}, 96s &\textbf{40\%}, 23\% &\begin{tabular}[c]{@{}l@{}}\textbf{60\%}, 24\%\end{tabular} &\textbf{60\%}, 24\% &\textbf{18\%}, 19\% \\\cline{2-8}
& \begin{tabular}[c]{@{}l@{}}4: Add an\\emp-pref\\message\end{tabular}           
& \begin{tabular}[c]{@{}l@{}}\textbf{68\%}, 31\%  \end{tabular}& \textbf{141s}, 76s &\textbf{36\%}, 20\% &\begin{tabular}[c]{@{}l@{}}\textbf{76\%}, 8\%\end{tabular} &\textbf{56\%}, 27\% &\textbf{24\%}, 19\% \\
\hline
{ \begin{tabular}[c]{@{}l@{}}Non-t\\messages\end{tabular}} 
&\begin{tabular}[c]{@{}l@{}}\ \\ \ \\ \ \end{tabular} 
&&&&& \textbf{37\%}, 25\%&\textbf{13\%}, 14\% \\
\hline
\multirow{5}{*}{ \begin{tabular}[c]{@{}l@{}}B:\\Top\\news\end{tabular}}    
& \begin{tabular}[c]{@{}l@{}}1: Original \\top news\end{tabular}          
& \begin{tabular}[c]{@{}l@{}}\textbf{68\%}, 38\% \end{tabular}&\textbf{146s}, 98s &\textbf{38\%}, 25\% & &\textbf{44\%}, 27\% &\textbf{17\%}, 15\% \\
\cline{2-8}
& \begin{tabular}[c]{@{}l@{}}2: Put\\random \\messages\end{tabular}          
& \begin{tabular}[c]{@{}l@{}}\textbf{58\%}, 29\% \end{tabular} &\textbf{123s}, 68s &\textbf{30\%}, 17\% & &\textbf{32\%}, 16\% &\textbf{9\%}, 12\% \\
\cline{2-8}
& \begin{tabular}[c]{@{}l@{}}3: Put\\org-pref \\messages\end{tabular}          
& \begin{tabular}[c]{@{}l@{}}\textbf{63\%}, 25\% \end{tabular} &\textbf{158s}, 81s &\textbf{46\%}, 23\% & &\textbf{49\%}, 16\% &\textbf{18\%}, 12\% \\
\cline{2-8}
& \begin{tabular}[c]{@{}l@{}}4: Put \\emp-pref \\messages\end{tabular}          
& \begin{tabular}[c]{@{}l@{}}\textbf{76\%}, 21\% \end{tabular} &\textbf{142s}, 82s &\textbf{37\%}, 21\% & &\textbf{49\%}, 21\% &\textbf{22\%}, 18\% \\
\cline{2-8}
& \begin{tabular}[c]{@{}l@{}}5: Mix \\emp-pref\\/org-pref\\messages\end{tabular} 
&\begin{tabular}[c]{@{}l@{}}\textbf{72\%}, 23\% \end{tabular} &\textbf{162s}, 123s &\textbf{49\%}, 19\% & &\begin{tabular}[c]{@{}l@{}}\textbf{53\%}, 17\%\\  \end{tabular}  &\textbf{18\%}, 14\% \\
\hline
\multirow{5}{*}{ \begin{tabular}[c]{@{}l@{}}C:\\Message\\order\end{tabular}}   
& \begin{tabular}[c]{@{}l@{}}1: Original \\message\\order\end{tabular}  
&\begin{tabular}[c]{@{}l@{}}\textbf{66\%}, 24\% \end{tabular} &\textbf{136s}, 66s &\textbf{40\%}, 23\% & & & \\
\cline{2-8}
& \begin{tabular}[c]{@{}l@{}}2: Sort \\messages\\randomly\end{tabular}  
&\begin{tabular}[c]{@{}l@{}}\textbf{69\%}, 28\% \end{tabular} &\textbf{127s}, 77s &\textbf{34\%}, 20\% & & & \\ 
\cline{2-8}
& \begin{tabular}[c]{@{}l@{}}3: Sort \\messages by\\org-pref\end{tabular}  
&\begin{tabular}[c]{@{}l@{}}\textbf{59\%}, 30\% \end{tabular} &\textbf{166s}, 124s &\textbf{46\%}, 25\% & & & \\
\cline{2-8}
& \begin{tabular}[c]{@{}l@{}}4: Sort \\messages by\\emp-pref\end{tabular}  
&\begin{tabular}[c]{@{}l@{}}\textbf{74\%}, 31\% \end{tabular} &\textbf{146s}, 84s &\textbf{41\%}, 20\% & & & \\ 
\cline{2-8}
& \begin{tabular}[c]{@{}l@{}}5: Sort \\messages\\by mix-order\end{tabular}      
&\begin{tabular}[c]{@{}l@{}}\textbf{74\%}, 19\%\end{tabular} &\textbf{157s}, 104s &\textbf{39\%}, 22\% & & & \\
\hline
\end{tabular}
}
\end{table}

\begin{table}[!htbp]
\caption{Results of hypotheses and research questions. The bold texts show (marginally) significance. \\ Format: experimental group mean = control group and its mean + difference between experimental and control groups (p.val). Blanks: not applicable or not of interest. Signif. codes:  `*' 0.05, `+' 0.1, `NS' no significant effect was found. 
Control groups: \\ rnd: random control group; org: original control group; \\
rnd-s: random subject lines' messages;  non-s: messages not on subject lines;\\
rnd-t: random top news' messages; org-t: original top news' messages; non-t: messages not in top news.\\
\hlc[white]{P-values were adjusted by Holm-Bonferroni correction for controlling Type I error} \cite{holm1979simple}.}~\label{tab:result}
\centering
\arrayrulecolor{black}
\resizebox{\textwidth}{!}{%
\begin{tabular}{|l|l|l|l|l|l!{\vrule width \heavyrulewidth}l|l|} 
\hline
                                  &                                                                                & \multicolumn{4}{c!{\vrule width \heavyrulewidth}}{\textbf{Strategic Goal}}                                                                                                                                                                                                                                                                                                                                                                                                                                                                              & \multicolumn{2}{c|}{\textbf{Tactical Goal}}                                                                                                                                                                                                                                                                                                \\ 
\hline
\begin{tabular}[c]{@{}l@{}}\textbf{ Group }\end{tabular}                  & \textbf{Treatment}                                                             & \begin{tabular}[c]{@{}l@{}}Interest Rate \\of Brief (\%) \end{tabular}                                                               & \begin{tabular}[c]{@{}l@{}}Reading Time of \\Brief (seconds)\end{tabular}                                                          & \begin{tabular}[c]{@{}l@{}}Recognition Rate \\of Brief (\%)\end{tabular}                                                                & \begin{tabular}[c]{@{}l@{}}Open Rate of\\Brief (\%)\end{tabular}                                          & \begin{tabular}[c]{@{}l@{}}Recognition Rate \\of Message (\%)\end{tabular}                                                                                                                   & \begin{tabular}[c]{@{}l@{}}Read-in-detail \\Rate of Message (\%)\end{tabular}                                                \\ 
\toprule

\multirow{3}{*}{ \begin{tabular}[c]{@{}l@{}}A:\\Subject\\lines\end{tabular}} 
& Anova P.val & (0.867) &(0.553) &(0.628) & (0.349)&\textbf{(0.001*)} &\textbf{(0.009*)}\\\cline{2-8}
&\begin{tabular}[c]{@{}l@{}}3: Add an \\org-pref \\message\end{tabular}         &   NS                                                                                                                                         &         NS                                                                                                                              &       NS                                                                                                                                       &             NS                                                                                            & \begin{tabular}[c]{@{}l@{}}H1.5 Increase  \\recognition rate.\\\textbf{60\%=non-s(41)+19 }\\\textbf{(0.001*)}\\\textbf{60\%=rnd-s(40)+20 }\\\textbf{(0.007*)}\end{tabular}                                                                     &   NS                                                             \\ 
\cline{2-8}
                                  & \begin{tabular}[c]{@{}l@{}}4: Add an\\emp-pref\\message\end{tabular}           & \begin{tabular}[c]{@{}l@{}}H1.1 Increase \\interest rate.\\ NS\end{tabular}                     & \begin{tabular}[c]{@{}l@{}}H1.2 Increase \\reading time.\\NS \end{tabular}                     & \begin{tabular}[c]{@{}l@{}}H1.3 Increase\\ overall\\recognition rate.\\NS \end{tabular}                        & \begin{tabular}[c]{@{}l@{}}H1.4 Increase \\open rate.\\NS\end{tabular} & \begin{tabular}[c]{@{}l@{}}H1.5 Increase  \\recognition rate.\\\textbf{56\%=non-s(41)+15 }\\\textbf{(0.001*)}\\\textbf{56\%=rnd-s(40)+16 }\\\textbf{(0.026*)}\end{tabular}                                                                    & \begin{tabular}[c]{@{}l@{}}H1.6 Increase  \\read-in-detail rate.\\\textbf{24\%=non-s(15)+9}\\\textbf{(0.008*)}\\rnd-s: NS\end{tabular}      \\ 
\hline
\multirow{4}{*}{ \begin{tabular}[c]{@{}l@{}}B:\\Top\\news\end{tabular}}    &Anova P.val&(0.182) &(0.809) &\textbf{(0.008*)} & &\textbf{(0.001*)} & \textbf{(0.001*)}\\\cline{2-8}  & \begin{tabular}[c]{@{}l@{}}3: Put\\org-pref \\messages\end{tabular}            &       NS                                                                                                                                  &    \begin{tabular}[c]{@{}l@{}}Q2.2 How does\\it affect \\reading time?\\ NS \end{tabular}                                                                                                                                       &        \begin{tabular}[c]{@{}l@{}}Q2.3 How does it\\affect overall\\recognition rate?\\org: NS\\\textbf{46\%=rnd(30)+16 }\\\textbf{(0.000*)}   \end{tabular}                                                                                                                                   &                                                                                                                & \begin{tabular}[c]{@{}l@{}}H2.5 Increase  \\recognition rate.\\\textbf{49\%=non-t(37)+12}\\\textbf{(0.002*)}\\rnd-t: NS\\org-t: NS\end{tabular}                                                                                         &NS                                                                 \\ 
\cline{2-8}
                                  & \begin{tabular}[c]{@{}l@{}}4: Put \\emp-pref \\messages\end{tabular}           & \begin{tabular}[c]{@{}l@{}}H2.1 Increase \\ interest rate.\\org:NS\\\textbf{76\%=rnd(58)+18}\\\textbf{ (0.075+)}\end{tabular} & \begin{tabular}[c]{@{}l@{}}Q2.2 How does\\it affect\\reading time? \\NS \end{tabular}                               & \begin{tabular}[c]{@{}l@{}}Q2.3 How does\\it affect overall\\recognition rate?\\NS \\\end{tabular}                       &                                                                                                                & \begin{tabular}[c]{@{}l@{}}H2.5 Increase  \\recognition rate.\\\textbf{49\%=non-t(37)+12} \\\textbf{(0.001*)}\\\textbf{49\%=rnd-t(32)+17} \\\textbf{(0.008*)}\\org-t: NS\end{tabular}                                                                  & \begin{tabular}[c]{@{}l@{}}H2.6 Increase  \\read-in-detail rate.\\\textbf{22\%=non-t(13)+9} \\\textbf{(0.001*)}\\\textbf{22\%=rnd-t(9)+13} \\\textbf{(0.005*)}\\org-t: NS\end{tabular}  \\ 
\cline{2-8}
                                  & \begin{tabular}[c]{@{}l@{}}5: Mix \\emp-pref\\/org-pref\\messages\end{tabular} &    NS                                                                                                                                          &  \begin{tabular}[c]{@{}l@{}}Q2.2 How does\\it affect \\reading time? \\NS \end{tabular}                                                                                                                                         &          \begin{tabular}[c]{@{}l@{}}Q2.3 How does it\\affect overall\\recognition rate?\\\textbf{49\%=org(38)+11}\\ \textbf{(0.069+)}\\\textbf{49\%=rnd(30)+19}\\\textbf{ (0.002*) } \end{tabular}                                                                                                                                  &                                                                                                                & \begin{tabular}[c]{@{}l@{}}H2.4 Increase  \\recognition rate of \\org-pref messages.\\orn\_B3: NS \\H2.5 Increase  \\recognition rate.\\\textbf{53\%=non-t(37)+16}\\\textbf{(0.001*)}\\\textbf{53\%=rnd-t(32)+21}\\\textbf{(0.050+)}\\org-t: NS\end{tabular} & NS                                                     \\ 
\hline
\multirow{4}{*}{ \begin{tabular}[c]{@{}l@{}}C:\\Message\\order\end{tabular}}    &  Anova P.val& \textbf{(0.088+)}&(0.674) &(0.446) & & &\\\cline{2-8}    & \begin{tabular}[c]{@{}l@{}}3: Sort \\messages by\\org-preference\end{tabular}  &   NS                                                                                                                                           & \begin{tabular}[c]{@{}l@{}}Q3.1 How does\\it affect\\reading time?\\NS\end{tabular}                      & \begin{tabular}[c]{@{}l@{}}Q3.2 Increase/\\decrease overall\\recognition rate?\\NS\end{tabular}            &                                                                                                                &                                                                                                                                                                                                   &                                                                                                                                     \\ 
\cline{2-8}
                                  & \begin{tabular}[c]{@{}l@{}}4: Sort \\messages by\\emp-preference\end{tabular}  & \begin{tabular}[c]{@{}l@{}}H3.1 Increase \\ interest rate.\\NS\end{tabular}                     & \begin{tabular}[c]{@{}l@{}}Q3.1 How does\\it affect \\reading time?\\NS\end{tabular}                     & \begin{tabular}[c]{@{}l@{}}Q3.2 How does it\\affect overall\\recognition rate?\\NS \end{tabular}             &                                                                                                                &                                                                                                                                                                                                   &                                                                                                                                     \\ 
\cline{2-8}
                                  & \begin{tabular}[c]{@{}l@{}}5: Sort \\messages\\by mix-order\end{tabular}       &       NS                                                                                                                                 & \begin{tabular}[c]{@{}l@{}}Q3.1 How does\\it affect \\reading time?\\NS \end{tabular} & \begin{tabular}[c]{@{}l@{}}Q3.2 How does it\\affect overall\\recognition rate?\\NS \end{tabular} &                                                                                                                & \begin{tabular}[c]{@{}l@{}}Q3.3 How does it\\affect recognition rate\\of org-pref messages?\\NS \end{tabular}                                                      &                                                                                                                                     \\
\hline
\end{tabular}
}
\end{table}

For the mixed logistic models in this paper, we checked \cite{kassambara2018logistic} 1) whether the numeric independent variables were linearly associated with the dependent variable in logit scale by visually plotting the line of predictor's value - logit of predicted probabilities; 2) multicollinearity (whether GVIF < 2);  3) whether outlier exists (by package dharma \cite{hartig2019dharma}). For the mixed linear model on reading time, we transformed $reading\_time$ and $base\_reading\_time$ by $log10(1+x)$. 3 outliers (the 3 emails were read for more than 30 minutes) were removed. The transformation and outlier removing were needed to satisfy the normality requirement of the model's residuals. We checked the homogeneity of variance by Levene’s Test and checked the normality of residuals by QQPlot \cite{roiger2020just}.

For each model and effect, first, we calculated the average of that performance metric of each experimental group. Second, we checked whether the effect was (marginally) significant by its ANOVA table to see whether there existed significant differences among different experimental groups (treatments) of this effect. If it was significant and we did observe large differences between some experimental groups with the control groups, we conducted pairwise tests between these experimental groups with its control groups (see Table \ref{tab:result} for the effects, treatments, and control groups we examined). The P-values of the pairwise tests were adjusted by the holm-bonferroni method \cite{holm1979simple}. We got the following marginal / significant results on the hypotheses and questions (see Table \ref{tab:result} for the numbers): 


 \noindent   \hlc[white]{\textbf{Interest rate}: 
H2.1 Putting employee-preferred messages in top news marginally increased Brief's interest rate versus putting random messages.}
    
\noindent \hlc[white]{\textbf{Overall recognition rate}:
     Q2.3 Mixing organization/employee-preferred  messages in top news increased Brief's overall recognition rate significantly versus putting random messages, marginally versus putting original messages. }
    
    \noindent \hlc[white]{Putting organization-preferred messages in top news significantly increased Brief's overall recognition rate versus putting random messages.}
    
    \noindent \hlc[white]{\textbf{Recognition rate}:
     H1.5 Putting messages on subject lines significantly increased their recognition rates versus the messages not on subject lines or putting random messages.}
    
    \noindent\hlc[white]{H2.5 Putting messages in top news significantly increased their recognition rates versus the messages not in top news.}

    \noindent \hlc[white]{\textbf{Read-in-detail rate}:
    H1.6 Putting employee-preferred messages on subject lines significantly increased their read-in-detail rates versus the messages not on subject lines.}
    
    \noindent\hlc[white]{H2.6  Putting employee-preferred messages in top news significantly increased their read-in-detail rates versus the messages not in top news or putting random messages.
    }

\subsection{Strategic Goals}
\hlc[white]{\textbf{Interest Rate}. Strategically, we could marginally make employees perceive Brief as more interesting by personalizing top news with their preferred messages. }We plot the average of the employees’ $interest\_rate$ on personalized Briefs of each
experimental group on top news in Figure \ref{fig:top4}. It shows that the average $interest\_rate$ of the employee-preferred top news group (B4) was higher than the random control group of top news (B2) by 18\%. The pairwise tests showed that \textbf{H2.1 was marginally supported versus the random control group}. In the end survey, a participant from B4 commented \textit{``I like the headings or topics at the top''}. A participant from B4 recognized that the top messages were related to their answers in the preference survey and hoped that recipients could update their preferences in the system in the future: \textit{``I liked having focused information. However, I think if you move forward with customized Briefs (which I support), people should get some what regular reminders with the ability to change what items they select to follow.''} A participant from the random group B2 seems to be disappointed: \textit{``I still think there's too much boosterism and fluff and I wish it was more work-related.''} And a participant from group B3 (which prioritized organization-preferred messages) found the content boring: \textit{``Most of it was skimmed. Most of the topics don't apply to me and/or my work. The content overall is generally uninteresting.''} The subject lines and message order's effects on the interest rate are not significant.

\noindent\hlc[white]{\textbf{Overall Recognition Rate}. However, putting employee-preferred messages in top news seemed to be a bad choice for the overall recognition rate} (see Figure \ref{fig:top_recog}, B4). The employees might close the Briefs early if they learned that most of the interesting messages would be at the top positions. As a participant from B4 said \textit{``It was fun to see the things I was interested in at the top. It also let me pay more attention to the beginning of the briefs and then skim the rest. ''}

\hlc[white]{To improve Brief's overall recognition rate, we could mix employee-preferred messages with organization-preferred messages in top news.} With pairwise tests, we found that the overall recognition rates of group B5 (mixing employee/organization preferences) were significantly higher than group B2 (random top news) by 19\% and marginally higher than group B1 (original top news) by 11\% (\textbf{Q2.3}). Seeing interesting content both at the top and in other sections might keep employees reading, though they did feel some of the messages ``irrelevant to them''. A participant from B5 said \textit{``I liked them. Overall I find things interesting; however, they are not really pertinent to my work always.''} It is worth noting that the overall recognition rate of the organization-preferred message group was also significantly higher than the random control group. The employees seemed to keep searching for items of interest if they did not find them in top news. But this searching process might cause disappointment. A participant from B3 commented \textit{``I was a little disappointed because I was expected slightly more tailored content.''} 
\begin{table}[!htbp]
\begin{minipage}{.49\textwidth}
  \centering
  \includegraphics[width=0.8\linewidth]{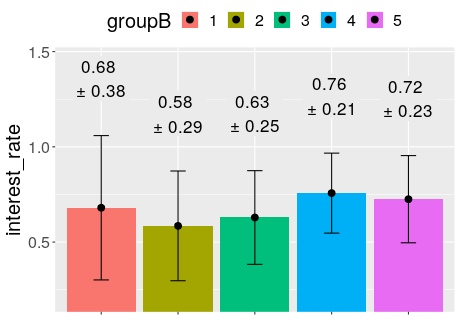}
  \captionof{figure}{The  interest rates on personalized Briefs of each experimental group on top news and the standard deviations of the participants'  interest rates in that group.}
  \label{fig:top4}
\end{minipage}%
\hspace{0.05in}
\begin{minipage}{.49\textwidth}
  \centering
  \includegraphics[width=0.85\linewidth]{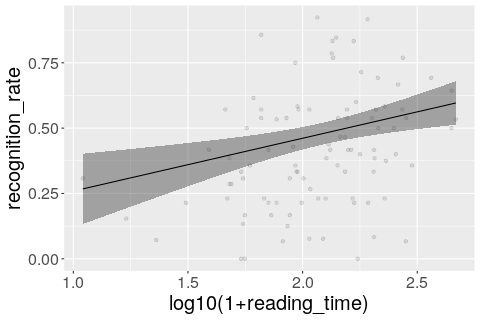}
  \captionof{figure}{Brief's reading time versus its  recognition rate.}
  \label{fig:time_recog}
\end{minipage}%

\end{table}

\begin{table}[!htbp]
\begin{minipage}{.49\textwidth}
  \centering
  \includegraphics[width=0.8\linewidth]{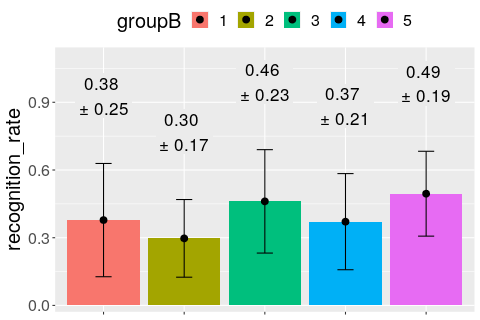}
  \captionof{figure}{The overall recognition rates on personalized Briefs of each experimental group on top news  and the standard deviations of the participants' overall recognition rates in that group.}
  \label{fig:top_recog}
\end{minipage}%
\hspace{0.05in}
\begin{minipage}{.49\textwidth}
  \centering
  \includegraphics[width=0.8\linewidth]{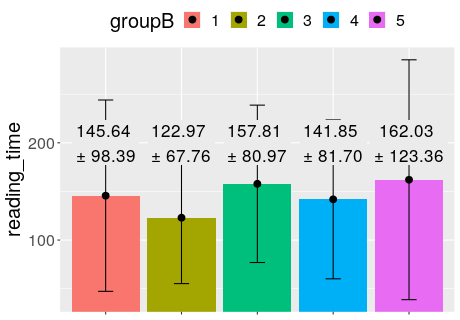}
  \captionof{figure}{The reading time (seconds) on personalized Briefs of each experimental group on top news  and the standard deviations of the participants' reading time in that group.}
  \label{fig:top_time}
\end{minipage}%
\end{table}

\noindent\textbf{Reading Time}. There were no significant effects of subject lines/top news/message order designs on reading time. On average, the Briefs were read for around 120 to 170 seconds, and the variation of reading time is large. However, we found that the patterns of reading time matched with the patterns of overall recognition rates (see \ref{fig:top_recog} and \ref{fig:top_time}). We plot reading time versus overall recognition rate in Figure \ref{fig:time_recog}. The correlation between reading time (transformed by log10(1+10)) and the overall recognition rate was significant (Chisq=10.46, p.value = 0.001,  coef = 0.20\textpm0.06). This result shows that the gain in awareness is usually accompanied by time costs.

\noindent\textbf{Open Rate}. We did not observe significant differences among subject line groups' open rates. The average $open\_rate$ of group A4 (employee-preferred subject line group) was higher than other subject line groups but the pairwise tests were insignificant. The reason might be that our participants usually would take a quick check of these experimental Briefs during the experiment (though we asked them to treat these Briefs as naturally as possible). This is a limitation of this study --- we only collected 6 weeks' datapoints, because we wanted to collect all the recognition data together by a survey with a reasonable amount of questions at the end. A longer study might find different results on the open rates. However, some participants did indicate that they decided whether to open a Brief or not based on its subject lines --- a participant from A2 said that they left two Briefs unread because their subject lines were ``not at all interesting''.

\subsection{Tactical Goals}
\noindent\hlc[white]{\textbf{Recognition Rate}. Tactically, organizations could make those messages they view as important/relevant be recognized by more employees by putting them on subject lines or top news.} The Anova tests showed that whether a message was on subject lines influences its recognition rate significantly. The pairwise tests showed that either putting organization-preferred messages or employee-preferred messages on subject lines would increase their recognition rates by over 15\% compared to the recognition rates of the messages not on subject lines (see Figure \ref{fig:subonly}). Similarly, either putting organization-preferred messages or employee-preferred messages in top news would increase their recognition rates by over 12\% compared to the recognition rates of the messages not in top news (see Figure \ref{fig:toponly}). It is worth noting that these recognition rates were not significantly higher than the recognition rates of the original top news groups (B1), which indicated an opportunity to learn from the human editor on their design and selection strategies.

\begin{table}[!htbp]
\begin{minipage}{.48\textwidth}
  \centering
  \includegraphics[width=0.75\linewidth]{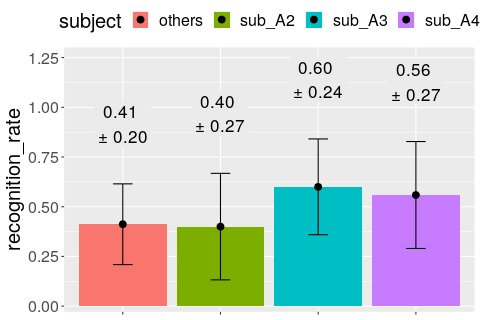}
  \captionof{figure}{The recognition rates of the messages on subject lines with respect to group A and non-subject line messages and the standard deviations of the participants'  recognition rates in that group.}
  \label{fig:subonly}
\end{minipage}
\hspace{0.05in}
\begin{minipage}{.48\textwidth}
  \centering
  \includegraphics[width=0.75\linewidth]{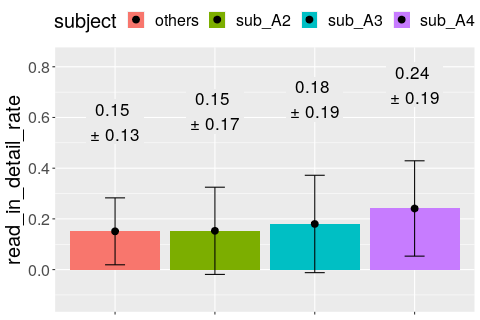}
  \captionof{figure}{The read-in-detail rates of the messages on subject lines with respect to group A and non-subject messages and the standard deviations of the participants'  read-in-detail rates in that group.}
  \label{fig:subread}
\end{minipage}
\end{table}

\noindent\hlc[white]{\textbf{Read-in-Detail Rate}. However, organizations could not make employees read the organization-preferred messages in detail.} The read-in-detail rates of those organization-preferred messages on subject lines were not significantly improved (see Figure \ref{fig:subread}). Actually, only the read-in-detail rates of those employee-preferred messages were significantly increased by 9\% when being putting on subject lines or top news (see Figure \ref{fig:subread}, \ref{fig:topread}). The reasons might be that the employees tended to only click the messages they had some interest in. We might need stronger incentives if we would like the employees to read those important-to-organization messages thoroughly.
\begin{table}[!htbp]
\begin{minipage}{.48\textwidth}
  \centering
  \includegraphics[width=0.75\linewidth]{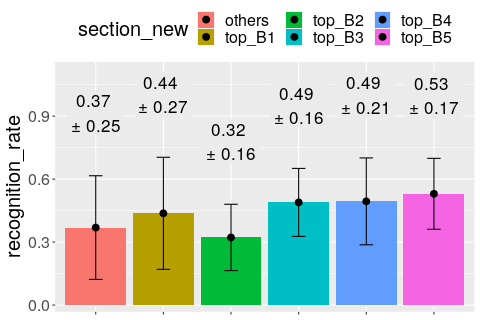}
  \captionof{figure}{The recognition rates of the messages in top news with respect to group B and non-top messages and the standard deviations of the participants'  recognition rates in that group.}
  \label{fig:toponly}
\end{minipage}
\hspace{0.05in}
\begin{minipage}{.48\textwidth}
  \centering
  \includegraphics[width=0.75\linewidth]{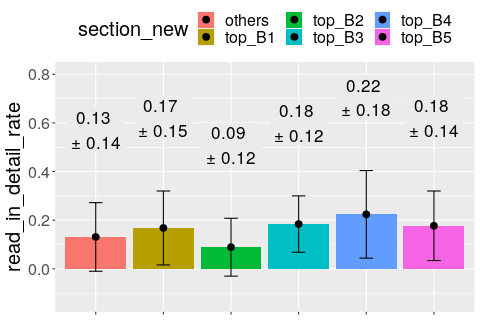}
  \captionof{figure}{The read-in-detail rates of the messages in top news with respect to group B and non-top messages and the standard deviations of the participants'  read-in-detail rates in that group.}
  \label{fig:topread}
\end{minipage}
\end{table}

Mixing organization-preferred messages with employee-preferred messages in top news/other sections did not bring further improvements to their recognition rates. For the messages in top news, though the mixed group B5's organization-preferred messages' average recognition rate was 4\% higher than B3's messages in the corresponding positions (45\% versus 49\%), the difference was not significant. And the difference (3\%) between the recognition rates of group C3 and C5's top 2 organization-preferred messages in the u-wide news sections was also not significant.

\subsection{Organization and Employee's Bulk Message Preferences}
In this section, we discussed where are the preference conflicts on bulk messages between the organization and the employees. Table \ref{tab:conflict} shows a set of messages' topics. For each topic, it shows how the Brief editor labeled the messages in that topic (whether it was important to the organization) in the editor surveys, and the employees' assessments of the work-relevance and interest of a sample message representing that topic (we collected these data in the preference survey at the experiment setup). For example, for the topic fundraising \& development, there were 9 messages during the study period. 5 of them were marked by the editor as important. 45.1\% of the employees felt the corresponding message was relevant to their jobs (18.0\% of the employees felt interesting and 38.5\% of them felt work-relevant).

We arranged a meeting with the Brief editor to discuss Table \ref{tab:conflict}. In that meeting, the editor told us that the frequency of topics (\#messages) is basically a true reflection of the number of topics submitted to them. The editor rejected a small number of the submissions that were too narrowly focused: \textit{``I really only reject maybe 10\% of submissions. We have communicators (in each campus). You know, it's really up to those folks to determine what they feel is important.''}. 


We noticed a number of interesting things from Table \ref{tab:conflict}. First, a large number of messages fell into the topic categories that the editor did not feel were usually important, and the employees generally would find unimportant and/or uninteresting, including award/recognition, student/alumni stories, faculty/staff stories. These were all work-relevant to fewer than 10\% of the employees and interesting to fewer than 20\% of them. 


Second, there are topics that the editor viewed as very important while the employees felt not that interesting or relevant, including university history \& celebrations, policy/admin news/governance, sports \& spirit. Over 60\% of the corresponding messages were viewed as important by the editor, while fewer than 40\% of the employees viewed these categories as work-relevant, and fewer than 30\% felt interesting.

Third, there are topics that the editor viewed as unimportant while the employees felt interesting. Some of them frequently appeared, including climate/eco, program awards/applications, health/covid. The editor told us that some contents were put because the employees might find them interesting: \textit{``I probably select half of those (the messages about events) myself just based on what I think folks will find interesting. I'm thinking both in terms of readership like we'd like them to find something of interest, so they come back and read.''} However, some of them appear only 3 or 4 times, including art \& museums and engineering science research stories. 

We further evaluated the preferences on the original Brief's messages. In week 2's Brief (the original Brief), 58\% of the surveyed messages were tagged as neither interesting nor work-relevant by the employees. The editor identified one message with the title ``U of M Public Engagement Footprint'' very relevant in building community and common understanding ($org\_importance = 4$). This message is from the Provost's office to advocate employees to submit plans for the university’s service, outreach, and community engagement. However, 58\% of the employees found this message neither interesting nor work-relevant. The message ``University and Faculty Senate Meetings'' was tagged as work-relevant to all employees, while 39\% of the employees found this message neither interesting nor work-relevant.

As we look at the results in total, it is clear that employee interest and editor judgment of importance are not perfectly-aligned. This finding reiterates the importance of considering the composition of the newsletter as a whole --- how to have enough relevant and interesting content to encourage reading the important content as well.
\begin{table}[!htbp]
\caption{Summary of the organization and employees' preferences on bulk messages' topics. The topics are ordered by
\# messages (the number of messages that included a corresponding topic in the experiment). The editor selected $\leq$ 4 topics for each message. \\
\# times\_imp: the number of times that the message got an importance score $\geq$ 3.\\ org\_imp\%: \# times\_imp / \# messages.\\
emp\_rel/int/pref\%: the percentage of employees who tagged the message with the corresponding topic as work-relevant/interesting/either work-relevant or interesting in the preference survey.
}~\label{tab:conflict}
\scalebox{0.8}{
\centering
\begin{tabular}{|p{4.3cm}|p{1.6cm}|p{1.7cm}|p{1.4cm}|p{1.5cm}|p{1.4cm}|p{1.4cm}|} 
\hline
\textbf{topic }                                                                                     & \textbf{\#messages } & \textbf{\#times\_imp } & \textbf{org\_imp\% } & \textbf{emp\_pref\% } & \textbf{emp\_rel\% } & \textbf{emp\_int\% }  \\ 
\hhline{|=======|}
Talk/ Symposium/ Lectures Announcements                    & 29                  & 1                      & 3.4\%                & 39.3\%                & 4.9\%                & 38.5\%                \\ 
\hline
Student/ Alumni Stories                                    & 27                  & 10                     & 37.0\%               & 18.0\%                & 4.9\%                & 14.8\%                \\ 
\hline
Community~Service/ Social Justice/ Underserved Population  & 21                  & 11                     & 52.4\%               & 78.7\%                & 35.2\%               & 73.0\%                \\ 
\hline
Faculty  Staff Stories                                     & 20                  & 4                      & 20.0\%               & 23.8\%                & 9.0\%                & 17.2\%                \\ 
\hline
Health/ Biology Research Stories                           & 15                  & 8                      & 53.3\%               & 64.8\%                & 9.0\%                & 60.7\%                \\ 
\hline
Climate/ Eco/ Agriculture                                  & 15                  & 6                      & 40.0\%               & 71.3\%                & 18.0\%               & 69.7\%                \\ 
\hline
Health Wellness Resources/ COVID                           & 12                  & 2                      & 16.7\%               & 91.0\%                & 67.2\%               & 73.8\%                \\ 
\hline
Award/ Recognition to University, Faculty, Staff, Students & 11                  & 5                      & 45.5\%               & 23.0\%                & 6.6\%                & 19.7\%                \\ 
\hline
Program  Award Applications/Announcements                  & 10                  & 2                      & 20.0\%               & 85.2\%                & 60.7\%               & 54.9\%                \\ 
\hline
Fundraising  Development                                   & 9                   & 5                      & 55.6\%               & 45.1\%                & 18.0\%               & 38.5\%                \\ 
\hline
History/ Social Science Research Stories                   & 9                   & 2                      & 22.2\%               & 45.9\%                & 15.6\%               & 38.5\%                \\ 
\hline
Policies/ Admin News/ Governance                           & 8                   & 5                      & 62.5\%               & 46.7\%                & 39.3\%               & 14.8\%                \\ 
\hline
Tech Tool Updates/ Workshops                               & 8                   & 0                      & 0.0\%                & 35.2\%                & 26.2\%               & 13.1\%                \\ 
\hline
Sports  Spirit                                             & 6                   & 5                      & 83.3\%               & 27.9\%                & 5.7\%                & 23.8\%                \\ 
\hline
University History/ Celebrations                           & 6                   & 4                      & 66.7\%               & 43.4\%                & 29.5\%               & 22.1\%                \\ 
\hline
Art  Museums                                               & 4                   & 0                      & 0.0\%                & 65.6\%                & 6.6\%                & 63.9\%                \\ 
\hline
University Program Success Stories                         & 4                   & 2                      & 50.0\%               & 39.3\%                & 17.2\%               & 27.9\%                \\ 
\hline
Operations Awareness/ Facility Closures                    & 3                   & 1                      & 33.3\%               & 89.3\%                & 82.0\%               & 49.2\%                \\ 
\hline
Engineering Science Research Stories                       & 3                   & 0                      & 0.0\%                & 54.1\%                & 3.3\%                & 52.5\%                \\ 
\hline
Youth, Children                                            & 0                   & 0                      & 0.0\%                & 36.1\%                & 8.2\%                & 31.1\%                \\
\hline
\end{tabular}}
\end{table}

\hlc[white]{For the engagement with topics versus campuses, only 25\%/32\% of the employees looked for work-relevant/interesting messages from other campuses. When being put in top news, the messages selected from other campuses got a recognition rate (21\%, p-value=0.0001) significantly lower than the messages selected from the employees' own campuses (43\%), and in this case, the employees' preference score $emp\_pref$ does not significantly influence these messages' recognition rate (this calculation has excluded the employees who indicated that they would not look at other campuses' messages). For the messages selected from the employees' own campuses, their recognition rate (43\%) is not significantly different from the messages originally selected from Top News (50\%), and in this case, the employees' preference score is positively correlated with the messages' recognition rate (p-value=0.024, cohen-size=2.259).}

\section{Discussion}
\hlc[white]{We explored 3 kinds of personalization (subject lines, top news, message order) based on 2 stakeholders' preferences (organization, employee), and investigated 2 types of goals (strategic / tactical goals). Overall, the tactical goals are easier to achieve than the strategic goals --- the organization could put whatever they want to promote in the top position and would get a reasonable recognition rate. But for the strategic goals, the organization needs to also consider the employees' information needs, and some strategic goals (reading time, open rate) can't achieve with blanket newsletters.}

\hlc[white]{Our work is different from many work in personalizing working emails \mbox{\cite{nelson2010mail2tag, moody2002reinventing, 10.1145/3290605.3300604}} in that we try to address the challenge of bulk emails in this multi-stakeholder case --- organizations have messages that they want their employees to be aware of while employees make individual judgements on which messages are relevant. Instead of prioritizing the recipients' preferences, we found that the best strategy for organizations is to mix messages they prefer with the messages their employees prefer. To the best of our knowledge, this is the first work focusing on this multi-objective personalization problem in the multi-stakeholder organizational bulk communication environment.}

\subsection{Organizations need to decide which messages to be sent and better communicate why.}
\hlc[white]{Organizations and employees perceive different messages as important/relevant --- this difference might have 2 outcomes. First, organizations might need to know more about their employees. For example, announcing the new Dean for the College of Biological Sciences through Brief is a convenient communication approach for the university leaders, because they do not need to spend time figuring out who would be interested in it and how to personalize the contents. However, many of its recipients would perceive this message as irrelevant, which might make them stop reading Brief in the future. In this case, organizations should collect more information to enable better targeting, such as collecting preferences based on message topics (this study), or allowing employees to select interesting message tags} \cite{nelson2010mail2tag}.

\hlc[white]{Second, if organizations decide that some messages are worthwhile for their employees to know about, they need to better communicate to their employees why they need to read those messages. For example, for the messages like ``Board of Regents Meeting Highlights'', the employees often skip reading them (62.5\% of the administrative news was viewed as important by the organization, while only 14.8\% of the investigated employees found this topic interesting). However, the employees might decide to read it if they are aware of, for example, that the board was discussing their salary plans. Potential approaches on this aspect include pricing emails \mbox{\cite{kraut2005pricing, reeves2008marketplace}} (however \mbox{\citeauthor{kraut2005pricing}} found that recipients still did not interpret the prices as emails' importance), indicating expected actions \mbox{\cite{alrashed2019evaluating, 10.1145/2556288.2557182}}, etc.  }


\subsection{Organizations could use employee-preferred messages to attract their attention.}
\hlc[white]{Though we found the only way to increase the recognition rates of those organization-preferred messages is to prioritize them, this does not mean that we can prioritize only these messages. Only prioritizing the organization-preferred messages would lose the gain in interest rate compared to putting employee-preferred messages on top, which might help protect this channel's long-term credibility --- 13\% more of the employees rated the Briefs with employee-preferred top news as interesting compared to the organization-preferred message group. Also, this choice might affect this Brief's open rate --- though insignificant, the open rate of the Briefs with organization-preferred subject lines is 16\% lower than the open rate of the employee-preferred subject line group.}

\hlc[white]{Only putting employee-preferred messages in top news might be a bad choice neither, though putting the contents interesting to its recipients is a common practice in commercial bulk emails. In our study, these Briefs' overall recognition rates were relatively lower (37\%) compared to putting organization-preferred messages in top news (46\%). The employees might be satisfied that they saw interesting messages early. However, this is undesirable for the organization because these employees might skip the rest of the newsletter, including the messages the organization wants them to know. In fact, when we mixed them in top news, the overall recognition rate would be the highest --- 11\% significantly higher than the original top news and 19\% higher than the random top news. This is, to some extent, similar to \mbox{\citeauthor{zhao2017toward}}'s finding that to keep users browsing a recommender website, we can not put all the interesting items at the top \mbox{\cite{zhao2017toward}}. Instead, we need to distribute it across the page and mix them with other items we would like to expose to users.}

\subsection{There are always tradeoffs --- suggestions.}
\hlc[white]{Within the current framework, we did not find any single optimal solution. Even with the mixed strategy, its interest rate was not as high as when we only put the messages the employees preferred on top news. The most interesting/efficient newsletter for employees would not be the newsletter that could best help organizations convey their messages. However, there are some decisions organizations can make when they know the priority of their communication goals:}

\noindent $\star$ \hlc[white]{Subject lines: for subject lines, organizations could put the messages they perceive as most important/relevant. This approach would bring these messages higher recognition rates. At the same time, at least in our (occasional) reader group, subject lines did not significantly affect the employees' open rate or interest rate.}

\noindent $\star$ \hlc[white]{Top news: to improve the overall recognition rate, a good approach for organizations would be mixing employee-preferred messages and organization-preferred messages in top news.}

Besides, there is also a trade-off between bulk communication's cost and performance. Though we did not find any significant effect on reading time, reading time is significantly correlated with the overall recognition rate, which means that organization needs to pay for more employee's time if they want higher performance. Also, longer email reading time is correlated with lower working productivity \cite{10.1145/2858036.2858262}. In that sense, organizations should also try to remove unnecessary messages and use personalization \cite{zhao2016group} to put important/relevant messages upfront.

\subsection{Limitations and Generalizability}
The limitations of this study included:
1) Reordering only: because of the requirement of our collaborator, we did not exclude any message from the studied newsletter.

\noindent2) Measurement of recognition: we trusted our participants that they would select ``Skimmed'' or ``Read fully'' if they have seen a message and select ``No'' or ``Not Sure'' if they did not recognize this message or were uncertain.

    
\noindent3) Selection of participants: our participants were relatively active readers of Briefs. The employees who had stopped reading Briefs might have lower recognition rates, open rates, etc.
    
\noindent4) Technical issues: some plugins were deleted by a Chrome update during the experiment, the participants did not read Briefs with that browser, etc. Our personalization model is based on coarse-grained topics. \footnote{Among the 1404 messages we sent through the original Briefs in week 2 (our predicted employee preference based on topics versus these messages’ employee preference calculated from the interest scores and work-relevance scores collected directly from the participants in the end survey, see 4.3), we achieve a precision of 66\% and a recall of 75\%.}
    
\hlc[white]{In short, our study could only be generalizable when: 1) the organization newsletter is sent to a large list of employees;}

\noindent \hlc[white]{2) studying newsletters' occasional/regular readers. We keep the newsletter's structure because the editor suggested that their audiences liked its campus structure; however, to how much extent the campus structure influenced the personalization's performance is still left to be studied.}

\subsection{Future Work}
After conducting this study, we see the following future work in improving organizational bulk communication. 

\noindent 1) Measuring each message's reading time to better understand employees' preferences \cite{10.1145/3588015.3588404}. It would be useful to run a study with eyetracking devices to collect such reading data, and to develop estimation algorithms \cite{kong2021nimblelearn} based on recipients' interactions with bulk emails' webpages.

\noindent 2) Exploring different designing strategies that could help employees understand why they need to read some messages: for example, encourage senders to tag the reasons for sending some messages \cite{he2023hiercat}.

\noindent 3) Studying the effectiveness of fine-grained personalization models \cite{aridor2022economics, sun2023less}, enabling employees to update their preferences, exploring tools could better target the recipients (allowing excluding some messages) and learn their preferences gradually \cite{shen2022classifying, yi2020natural}, etc.

\noindent 4) Studying how to bring back nonreaders: restore nonreaders' trust on the bulk communication channels.

\section{Conclusion}
This work studied how to use personalization to help the studied organization lead their employees' attention to the bulk messages they perceive as important or relevant for the employees to know (tactical goals) while maintaining the employees' overall positive experiences with these emails (strategic goals). We conducted an 8-week 4x5x5 controlled field experiment with 141 employees of a university and a weekly university-wide newsletter. 

\hlc[white]{We found that tactically, putting organization-preferred messages on subject lines or top news significantly increased their recognition rates, but did not increase their read-in-detail rates significantly. Only the employee-preferred messages' read-in-detail rates were improved. Strategically, mixing the employee-preferred/organization-preferred messages in top news significantly increased the overall recognition rate. Putting employee-preferred messages in top news increased their interest rates marginally. } We further looked into where the preferences on bulk messages' topics conflicted between the employees and the organization. We discussed the limitations and generalizability of this study.

Besides the findings above, this work also provided a basic backend framework for coordinating multiple stakeholders' preferences on organizational bulk emails --- employees could input their preferences through an onboarding survey; communicators and the organization leaders could input their preferences through weekly surveys; the system handles the transformation between text and html, listens to the survey responses, and generates personalized newsletters.

To the best of our knowledge, this is the first work focusing on this multi-objective personalization problem in the multi-stakeholder organizational bulk communication environment. We hope our study provides facts and possible directions for designing tools supporting organizational bulk communications.

\begin{acks}

  This work was supported by the National Science Foundation under grant CNS-2016397. We thank the University of Minnesota's communication professionals, Adam Overland, Benjamin Peck, and Kellie Greaves, for their assistance with bulk email design and communication.
\end{acks}

\bibliographystyle{ACM-Reference-Format}
\bibliography{sample-base}


\appendix
\begin{table}[!htbp]
\caption{The summary of bulk messages' topics of Brief and employees' job categories of the studied site.}
\centering
\arrayrulecolor[rgb]{0.8,0.8,0.8}
\scalebox{0.7}{
\begin{tabular}{!{\color{black}\vrule}l|l|l!{\color{black}\vrule}l!{\color{black}\vrule}} 
\arrayrulecolor{black}\hline
\multicolumn{3}{!{\color{black}\vrule}c!{\color{black}\vrule}}{\textbf{Appendix A: Message Topics }}                                                                                                                                                                       & \multicolumn{1}{c!{\color{black}\vrule}}{\textbf{Appendix B: Job Categories}}                                                                                           \\ 
\arrayrulecolor[rgb]{0.8,0.8,0.8}\hline
\begin{tabular}[c]{@{}l@{}}Talk/Symposium/\\Lectures Announcements\end{tabular}  & \begin{tabular}[c]{@{}l@{}}Community Service/ Social \\Justice/Underserved Population\end{tabular}  & Sports  Spirit                                                           & \begin{tabular}[c]{@{}l@{}}Administration \& Advancement \\\& Communication Staff\end{tabular}                                                                 \\ 
\hline
\begin{tabular}[c]{@{}l@{}}Operations Awareness\\/Facility Closures\end{tabular} & \begin{tabular}[c]{@{}l@{}}Award/Recognition to University, \\Faculty, Staff, Students\end{tabular} & Youth, Children                                                          & \begin{tabular}[c]{@{}l@{}}Campus Operation Staff (e.g., facilitates \\maintenance, dining services, police,\\~bookstore, athletics operations.)\end{tabular}  \\ 
\hline
\begin{tabular}[c]{@{}l@{}}Health Wellness \\Resources/COVID\end{tabular}        & \begin{tabular}[c]{@{}l@{}}Program \& Award Applications\\/Announcements\end{tabular}               & Art  Museums                                                             & \begin{tabular}[c]{@{}l@{}}Faculty, Teaching \& Research Staff, \\Librarians, Museum Curators \\and Directors, etc.\end{tabular}                               \\ 
\hline
\begin{tabular}[c]{@{}l@{}}Fundraising \& \\Development\end{tabular}             & \begin{tabular}[c]{@{}l@{}}University Program \\Success Stories\end{tabular}                        & \begin{tabular}[c]{@{}l@{}}Policies/Admin News\\/Governance\end{tabular} & \begin{tabular}[c]{@{}l@{}}Healthcare Staff (nurses, doctors, \\athletic trainers, etc.)\end{tabular}                                                          \\ 
\hline
\begin{tabular}[c]{@{}l@{}}Climate/Eco/\\Agriculture\end{tabular}                & \begin{tabular}[c]{@{}l@{}}Engineering Science \\Research Stories\end{tabular}                      & Faculty  Staff Stories                                                   & \begin{tabular}[c]{@{}l@{}}Human Resource \& Finance Staff \\(e.g., accountants, HR specialists)\end{tabular}                                                  \\ 
\hline
\begin{tabular}[c]{@{}l@{}}University History\\/Celebrations\end{tabular}        & \begin{tabular}[c]{@{}l@{}}History/Social Science \\Research Stories\end{tabular}                   & Student/Alumni Stories                                                   & Information Technology Staff                                                                                                                                   \\ 
\hline
\begin{tabular}[c]{@{}l@{}}Tech Tool Updates\\/Workshops\end{tabular}            & \begin{tabular}[c]{@{}l@{}}Health/Biology Research \\Stories\end{tabular}                           &                                                                          & \begin{tabular}[c]{@{}l@{}}Student Services Staff (advisors, \\student union staff, financial aid staff, etc.)\end{tabular}                                    \\
\arrayrulecolor{black}\hline
\end{tabular}
}
\end{table}

\received{January 2022}
\received[revised]{April 2022}
\received[accepted]{August 2022}
\end{document}